\documentclass{jfm}
\usepackage{graphicx}
\usepackage{epstopdf, epsfig}
\newcommand{\tabincell}[2]{\begin{tabular}{@{}#1@{}}#2\end{tabular}}

\usepackage{xcolor}
\shorttitle{Turbulent convection over rough plates with varying roughness geometries}
\shortauthor{Yi-Chao Xie and Ke-Qing Xia}

\title{Turbulent thermal convection over rough plates with varying roughness geometries}

\author{Yi-Chao Xie \and Ke-Qing Xia\corresp{\email{kxia@phy.cuhk.edu.hk}}}
\affiliation{Department of Physics, The Chinese University of Hong Kong, Shatin, Hong Kong, China}
\begin{document}
\maketitle
\begin{abstract}

We present a systematic investigation of the effects of roughness geometry on turbulent Rayleigh-B\'enard convection (RBC) over rough plates with pyramid-shaped and periodically distributed roughness elements. Using a parameter $\lambda$ defined as the height of a roughness element over its base width, the heat transport, the flow dynamics and local temperatures are measured for the Rayleigh number range $7.50\times 10^{7} \leq Ra\leq 1.31\times 10^{11}$, and the Prandtl number $Pr$ from 3.57 to 23.34 at four values of $\lambda$ (0.5, 1.0, 1.9, and 4.0). It is found that the heat transport scaling, i.e. $Nu\sim Ra^{\alpha}$ where $Nu$ is the Nusselt number, may be classified into three regimes in turbulent RBC over rough plates. In Regime I, the system is in a dynamically smooth state. The heat transport scaling is the same as that in a smooth cell. In Regimes II and III, the heat transport enhances. When $\lambda$ is increased from 0.5 to 4.0, $\alpha$ increases from 0.36 to 0.59 in Regime II, and it increases from 0.30 to 0.50 in Regime III. The experiment thus clearly demonstrates that the heat transport scaling in turbulent RBC can be manipulated using $\lambda$ in the heat transport enhanced regime. Previous studies suggest that the transition to heat transport enhanced regime, i.e. from Regime I to Regime II, occurs when the thermal boundary layer (BL) thickness becomes smaller than the roughness height. Direct measurements of the viscous BL in the present study suggest that the transition from Regime II to Regime III is likely a result of the viscous BL thickness becoming smaller than the roughness height. The scaling exponent of the Reynolds number $Re$ with respect to $Ra$ changes from 0.471 to 0.551 when $\lambda$ is increased from 0.5 to 4.0, suggesting a change of the dynamics of the large-scale circulation. Interestingly, the transition from Regime II to Regime III in terms of the heat transport scaling is not reflected in the $Re$-scaling with $Ra$. It is also found that increasing $\lambda$ increases the clustering of thermal plumes which effectively increases the plumes lifetime. This leads to a great increase in the probability of observing large temperature fluctuations in the bulk flow, which corresponds to the formation of more coherent plumes or plume clusters that are ultimately responsible for the enhanced heat transport.
\end{abstract}


\section{Introduction}

Turbulent thermal convection plays very important role in a range of phenomena in astrophysical and geophysical systems, where the underlying surfaces over which convective turbulence occurs are always rough. As an idealised model for the study of turbulent thermal convection in general, turbulent Rayleigh-B\'enard convection (RBC), where a fluid layer confined between two horizontally parallel plates heated from below and cooled from above, has been studied extensively in the past two decades \citep*[for reviews, see e.g.][]{Siggia1994ARFM,Ahlers2009RMP,Xia2010ARFM,SchumacherChilla2012,Xia2013TAML}. In terms of equations of motion, there are two control parameters in RBC, i.e. the Rayleigh number $Ra=\alpha g\Delta T H^3/(\nu\kappa)$ and the Prandtl number $Pr=\nu/\kappa$, where $ \Delta T$ is the applied temperature difference across a fluid layer of height $H$, $g$ is the gravitational acceleration constant and  $\alpha$, $\kappa$ and $\nu$ are respectively the thermal expansion coefficient, the thermal diffusivity and the kinematic viscosity of the working fluid. Any laboratory convection experiments are confined, thus the aspect ratio $\Gamma$ and the container shape come into the problem. One of the central issues in the study of turbulent RBC is to determine how the response parameters, i.e. the Nusselt number $Nu$ and the Reynolds number $Re$, depend on the control parameters, where $Nu$ is a measurement of the system's heat transport efficiency and $Re$ quantifies the turbulence intensity. The functional forms of $Nu(Ra, Pr)$ and $Re(Ra, Pr)$ are usually expressed in terms of power laws:
\begin{equation}
Nu\sim Ra^{\alpha}Pr^{\zeta} \qquad Re \sim Ra^{\beta}Pr^{\epsilon} 
\end{equation}
A comprehensive review on this issue can be found in \citet{Ahlers2009RMP}. It is now generally accepted that there are multiple scalings in the $Ra-Pr$ phase space where the bulk flow is in a turbulent state according to the Grossmann and Lohse model for turbulent thermal convection \citep{GL2000JFM,GL2001PRL,GL2002PRE,GL2004PoF}.

Besides its fundamental importance in geophysical and astrophysical problems, turbulent thermal convection over rough surfaces is found to be a more effective way to transport heat, which is also useful in practical applications. Heat transport enhancement is observed when the top and bottom plates are rough and the thermal boundary layer (BL) is thinner than the roughness height \citep*{Shen1997PRL}. The enhancement is later found to be due to more thermal plumes being emitted from the tip of the roughness elements \citep*{Du1998PRL}.

Although the heat transport is enhanced, the scaling relation between $Nu$ and $Ra$ is found to be insensitive to the plate morphology \citep{Shen1997PRL,Du1998PRL}. Several later experiments show that the scaling exponents also change in convection cells with rough plates (rough cells hereafter) when compared with observations in convection cells with smooth plates (smooth cells hereafter). Using glass spheres adhered to the bottom plate of the convection cell, the scaling exponent is found to depend on the distribution of the roughness height \citep{Ciliberto1999PRL}. For $Ra>10^{12}$, a scaling exponent of $0.5$ is reported when grooved roughness is distributed on both the plates and sidewall of the convection cell \citep{Roche2001PRE}. A change of the scaling exponent from 0.28 in a smooth cell to 0.35 in a rough cell is observed \citep*{Qiu2005JoT}. The rough cell used in the experiments of \citep{Qiu2005JoT} had similar design with the ones used in \citet{Shen1997PRL} and \citet{Du1998PRL}.These authors have suggested that the discrepancy between their experiments and those of \citet{Shen1997PRL} and \citet{Du1998PRL} may be attributed to the different constructions and materials of the rough plates. In the earlier experiments, rough plates made of brass were attached on the surfaces the top and bottom smooth plates. While in the later experiment, the roughness elements were directly machined on a piece of copper. In addition to studying the global transport properties of the convection cell, the heat transport properties of individual plates with rough/smooth surfaces were also investigated \citep{Wei2014JFM}. It is found that heat transport property of individual plate differs significantly and depends on the nature of the opposite plate of the same convection cell and the pertaining temperature boundary conditions. By determining the $Nu$ separately for the bottom rough plate and the top smooth plate, a scaling exponent of 1/2 for the rough plate and of 1/3 for the smooth plate are reported \citep{Tisserand2011PoF}.

The effects of roughness elements on the local dynamics of turbulent RBC are also studied. Using flow visualization and near-wall temperature measurements, it is shown that the interaction of the large-scale circulation (LSC) and the secondary flow in the grooved region of the rough plate enhances plume emission which leads to the enhancement of heat transport \citep{DU2000JFM}. The study of local temperature fluctuations suggest that the enhancement of heat transport is determined by the dynamics in the near plate region \citep{Du2001PRE}. The effects of rough plates on the velocity BL have been investigated very recently \citep{Liot2016JFM}, where it is found that the flow field is very different before and after the heat transport enhancement transition.  

Direct numerical simulations (DNSs) of turbulent RBC in rough cells also show that the scaling exponent increases \citep*{Verzicco2006JFM,Olga2011JFM,Wanger2015JFM,Wettlaufer2015EPL,Wettlaufer2017PRL}. The scaling exponent $\alpha$ increases to 0.37 in a rough cell with grooved walls at the top and bottom plates when compared with the smooth cell \citep{Verzicco2006JFM}. The scaling exponent was also found to depend on the geometry of the roughness elements or ``obstacles" in a quasi-2-D DNS \citep{Wanger2015JFM}. But the number of roughness elements in this study is very limited (only 4 ``obstacles" are used). In a 2-D study, using sinusoidal rough elements, the scaling exponent $\alpha$ is also found to depend on the wavelength of the roughness elements \citep{Wettlaufer2015EPL,Wettlaufer2017PRL}. In a DNS study of turbulent Taylor-Couette flow the scaling exponent is also found to increase with wall-roughness \citep{Zhu2016JFM}. These numerical studies imply the importance of roughness geometry on turbulent transport. However, the limited number of roughness element in a 3-D or a (quasi-) 2-D computation domain makes the comparison between the DNS studies and experimental ones less straightforward. 

Although there are quite a number of previous experiments studying the effect of roughness on turbulent RBC, to the best of our knowledge no experimental study has been made to systematically varying roughness geometry. This motivates the present study in which the effects of roughness geometry on turbulent RBC have been investigated by systematically varying the roughness geometry. A parameter $\lambda$ which is defined as the height of a single roughness element over its base width is used to characterise the geometry of roughness elements. In addition to heat transport enhancement as found by previous studies, we find that the heat transport scaling in a rough cell with fixed roughness height $h$ may be classified into three regimes. The experiments further show that $\lambda$ can be used as a tuning parameter to manipulate heat transport scaling in rough cells in the heat transport enhanced regime. The Reynolds number and local temperature fluctuation scalings with $Ra$ also change with $\lambda$, suggesting that the geometry of roughness elements not only alters the global heat transport of the system but also influences the bulk turbulence and the dynamics of the LSC.

The remainder of the paper is organised as follows: The experimental setup and measurement techniques are introduced in \S \ref{Sec_Setup} which is divided into four subsections. The construction of the convection cell is described in \S \ref{SubSec_cell}. The properties of the working fluids  used and the explored $Ra-Pr$ phase space are introduced in \S\ref{SubSec_fluid}. The temperature measurement techniques are documented in \S \ref{SubSec_temp}, and \S \ref{SubSec_PIV} introduces the measurement of the viscous BL using particle image velocimetry (PIV). \S \ref{Sec_Results} presents the experimental findings. The heat transport measurements for different values of $\lambda$ are reported in \S \ref{SubSec:Nu}. The transitions between different heat transport regimes are discussed in \S \ref{SubSec_Transition}. The Reynolds number and local temperature fluctuation measurements are presented in \S \ref{SubSec_Re} and \S \ref{SubSec_Temp_fluc}, respectively. A discussion of the proper parameter that characterises the system is reported in \S \ref{SubSec_con_para} before we summarise and conclude our findings in \S \ref{Sec_con}.

\section{Experimental setup and measurement techniques\label{Sec_Setup}}

\subsection{The convection cell\label{SubSec_cell}}

\begin{figure}
\includegraphics[width=\textwidth]{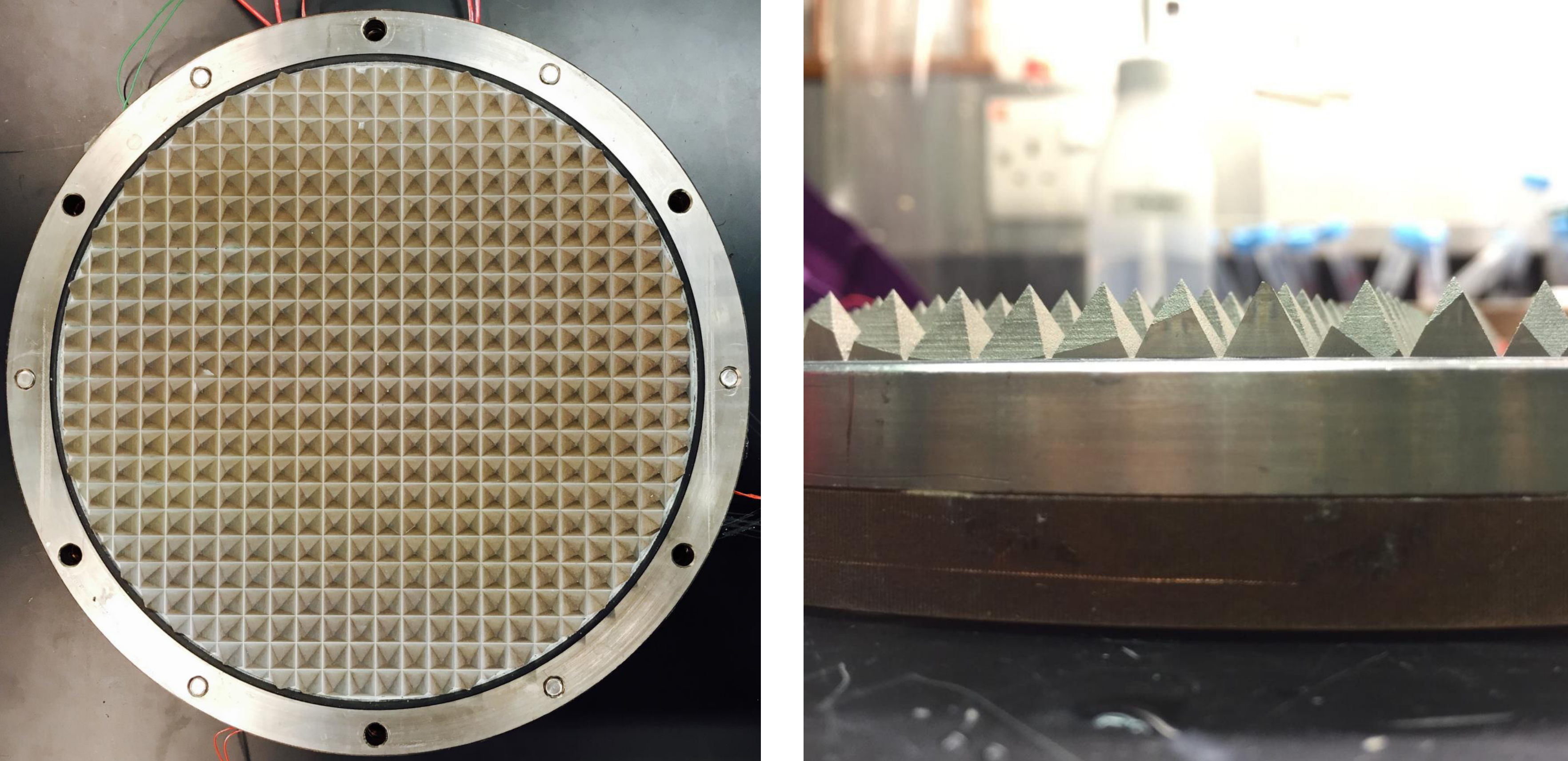}
\caption{\label{fig_pic}(Colour online) Photos of the bottom plate of a rough cell with $\lambda=1.0$. Left: top view; right: side view. The red heater connection wires can be seen on top of the left panel. Plates with $\lambda$ values of 0.5, 1.9 and 4.0 have the same design.}
\end{figure}

The experiments were carried out in four cylindrical convection cells. Each of the convection cells consisted of top and bottom plates and a Plexiglas sidewall. Except for the geometry of roughness elements on the top and bottom plates, the constructions of the convection cells were otherwise identical.

The sidewall was made of Plexiglas tube with a wall-thickness of 4 mm. The convection cells had a diameter $D$ of 19.2 cm and a height $H$ of 20.3 cm measured from the base of the roughness elements. Thus the aspect ratios $\Gamma=D/H$ of the convection cells were approximately unity. The top and bottom plates were made of 12 mm-thick copper except for the ones for rough cell with $\lambda=1.9$ which were made of aluminium. The thermal conductivity of aluminium is 205 W/(m$^o$C) at temperature 25 $^o$C, which is roughly half that of copper. The differences in heat transport properties in convection cells with plates made of copper and aluminium, i.e. the effects of finite conductivity of plates \citep{Verzicco2004PoF,Brown2005PoF}, have been shown to be minimal in the $Ra$ and $Pr$ range reported in the present studies \citep*{Wei2012PRE}. The pyramid-shaped roughness elements were directly machined on the top and bottom plates that were then plated with nickel to prevent oxidation. The roughness elements were arranged in a square lattice form (see figure \ref{fig_pic}), which was the same as those used previously \citep{Wei2014JFM, Xie2015JFM}. By keeping the height of roughness elements $h$ a constant at 8 mm and changing the base width $w$ of individual roughness elements from 16 mm to 2 mm, rough plates with roughness parameter $\lambda$, defined as the ratio of $h$ and $w$, at 0.5, 1.0, 1.9 and 4.0  were used in the experiments. With the increase of $\lambda$, the number density $n$ (in unit of number of roughness element per cm$^2$) also increases, i.e. $n$ = 0.38 ($\lambda = 0.5$), 1.5 ($\lambda = 1.0$), 6.1 ($\lambda = 1.9$), and 24.5 ($\lambda = 4.0$). To clarify whether the number density or the $\lambda$ is the key parameter that characterizes the system, another sets of rough plates with $\lambda$ = 4.0 and a roughness height of 4 mm were used. The present study implies that $\lambda$ is a more suitable parameter to characterise the geometry of the roughness elements and the response of the system, rather than the number density of the roughness elements. 

The top plate was cooled by passing temperature-controlled water from a water bath (Poly Science, 9702) through a chamber fitted on its top. The cooling water entered the chamber through two inlets and left the chamber through two outlets. The arrangements of the inlets and outlets were such that the two inlets were along one diameter of the plate and the two outlets were along another diameter perpendicular to the inlets. Temperature stability of the cooling water was better than 0.01 $^o$C. A stirring blade driven by the inlet and outlet flows was installed inside the cooling chamber to ensure temperature uniformity across the top plate. The bottom plate was electrically heated by two rubber heaters connected in series and sandwiched in between the bottom plate and a copper plate with a thickness of 10 mm (see, e.g. figure \ref{fig_pic}). The heaters were connected to a DC power supply with a long term voltage stability of 99.99\% (Sorensen, XFR300-4). To ensure good thermal contact, a thin layer of thermally conductive paste was used between the heaters and the plates. The largest temperature difference within the same plate was about $1.5 \% \Delta T$. It should be noted that under this arrangement, the temperature boundary condition for the top plate was constant temperature and that for the bottom plate was constant heat flux.

To prevent heat leakage and to minimise the influence of the environmental temperature fluctuation, the convection cells were wrapped with Styrofoam with a thickness of 6 cm. A temperature regulated copper basin was placed beneath the convection cell to prevent heat leakage from the bottom plate of the convection cell by maintaining the same temperature at the top and bottom of the basin. The whole set up was put into a thermostat where the temperature was regulated to match the mean temperature $T_m$ of the top and bottom plate. The temperature stability of the thermostat was better than 0.1 $^o$C.

\subsection{The working fluids \label{SubSec_fluid}}

Only about one and a half decades of $Ra$ can be achieved in a single experimental setup using classical working fluid like water. To extend the $Ra$ range, two kinds of working fluids, namely deionized  water and Flourinert FC770 (FC770 hereafter, 3M Inc.) were used. Water has been widely used in the study of turbulent convection, we will mention here only the properties of FC770. The data listed below are given by the manufacturer (3M Inc.) and are all obtained at the temperature $25^{o}$C. The FC770 has  $\kappa$ = 0.063 W/(m$^o$C), $\alpha$ = 0.00148/$^o$C, $\nu = 7.9\times 10^{-7} $ m$^2$/s, the specific heat $c_p = $ 1038 J/(kg$^o$C), and the density $\rho$ = 1793 kg/m$^3$. 

The working fluid was filled into the convection cell through a hole attached with a stainless steel tube ($\sim$ 8 mm in diameter) located at the centre of the top plate. The working fluids were degassed in the following way to ensure there was no bubble forming during a single run of the experiment, which typically lasts for three weeks. Deionized water was boiled for a hour and then cooled down to room temperature before filling into the convection cell. For FC770, it was first filled into the convection cell and degassed by keeping the temperature of the top and bottom plate, and thus the whole convection cell, at $50 ^o$C for 24 hours. Then the temperature of the convection cell was lowed to $25 ^o$C which was the $T_m$ for all the experiments using FC770, and the bubbles formed were totally removed.

\begin{figure}
\begin{center}
\includegraphics[width=0.65\textwidth]{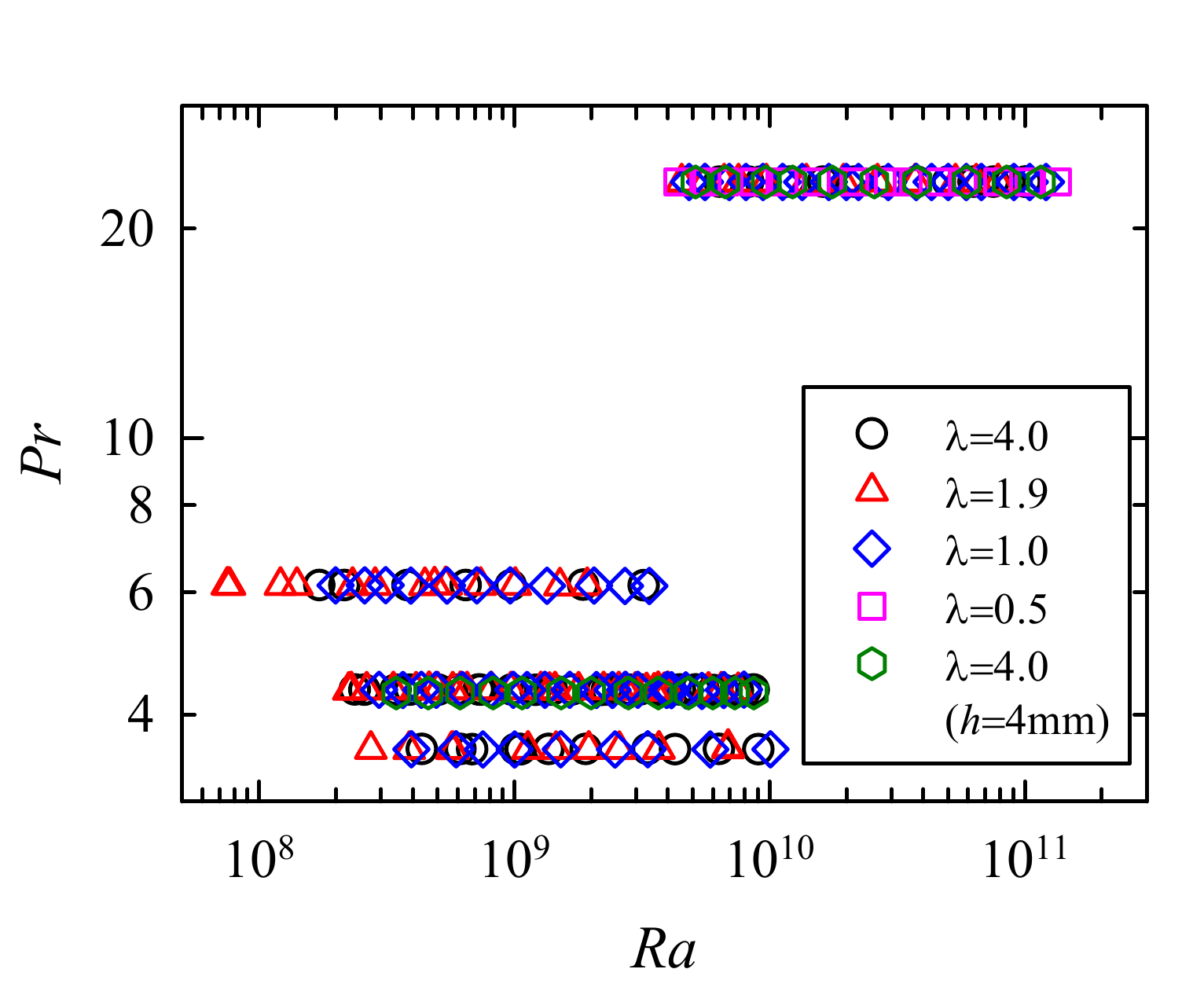}
\caption{\label{fig_PhaseSpace}(Colour online) $Ra-Pr$ phase space explored for different values of $\lambda$. The case for $\lambda$ = 0.5 is only studied for $Pr=$23.34.  All experiments were conducted in convection cells with roughness height $h$ = 8 mm, except for the ones corresponding to hexagons that were conducted in convection cell with $h$ = 4 mm.}
\end{center}
\end{figure}

The explored $Ra$ and $Pr$ phase space at different values of $\lambda$ is shown in figure \ref{fig_PhaseSpace}. Except for $\lambda=0.5$ which was investigated for a single $Pr$ (23.34) in the present studies, all the other values of $\lambda$ were studied for the four values of $Pr$, i.e. $Pr$ = 3.67 (water, $T_m = 50.00$  $^o$C), $Pr$ = 4.34 (water, $T_m = 40.00$ $^o$C), $Pr$ = 6.14 (water, $T_m = 25.00$ $^o$C), and $Pr$ = 23.34 (FC770, $T_m = 25.00 $ $^o$C). The experiments covered $7.50\times 10^7\leq Ra \leq 1.31\times 10^{11}$. The temperature difference applied across the convection cell using water as the working fluid for different values of $\lambda$ were: 0.98 $^o$C to 26.99 $^o$C ($\lambda = 4.0$); 0.47 $^o$C to 20.16 $^o$C ($\lambda = 1.9$) and 0.89 $^o$C to 24.75 $^o$C ($\lambda = 1.0$). The temperature difference applied using FC770 as working fluid at different values of $\lambda$ were: 1.40 $^o$C to 22.14 $^o$C ($\lambda = 4.0$); 0.99 $^o$C to 17.25 $^o$C ($\lambda = 1.9$); 1.11 $^o$C to 27.63 $^o$C ($\lambda = 1.0$) and 0.96 $^o$C to 29.66 $^o$C ($\lambda = 0.5$).

\subsection{The temperature measurements \label{SubSec_temp}}

Temperature of the top plate was measured using four thermistors (Omega, Model 44031) embedded at a distance of $D/4$ from the centre of the plate and uniformly in the azimuthal direction. Four thermistors at the mirror location of the top plate embedded in the bottom plate with one additional thermistor at the centre were used to measure its temperature. The thermistors were calibrated individually and separately in a temperature controlled water bath. The temperature range of calibration was from 5 $^o$C to 70 $^o$C with a temperature accuracy better than 0.01 $^o$C. The resistances of the thermistors were measured using a $6\frac{1}{2}$ digital multimeter (Keithley, 2700) and were converted to temperature using the calibration curve. The sampling rate of the temperature measurement was 0.7 Hz. 

Local temperatures at the cell centre and $\sim$ 1 cm away from the sidewall were measured using two small thermistors with a bead diameter of $0.38$ mm and a time constant of 30 ms in liquid (Measurement Specialities, Model: G22K7MCD419). These small thermistors had a temperature accuracy of 0.01 $^o$C. They were guided into the convection cell using a stainless steel tube with an outer diameter of 1.1 mm and wall thickness of 0.2 mm (Goodfellow, Stainless Steel - AISI 304 Tube). Each of the small thermistors had an independent temperature measurement system specially designed to measure the temperature fluctuation. Detailed design of the measurement system can be found in \citet{Zhou2001PRL}. Briefly, the thermistor acted as one arm of a AC Wheatstone electrical bridge. The resistance fluctuations of the thermistor induced by the temperature fluctuations in the convection cell were converted to the voltage fluctuations of the electrical bridge which were magnified using a lock-in amplifier (Stanford Research System, SR830). The output signal from the lock-in amplifier was then digitised and stored by a dynamic signal analyser (HP, 35670A). The measured voltage was converted to resistance and then to temperature using the calibration curve of the thermistors. The sampling rate of the local temperature varied from 16 Hz to 128 Hz depending on $Ra$ and $Pr$ to ensure that the small scale temporal temperature fluctuations were fully resolved. 

From the temperature measurements, the heat transport across the convection cell, i.e. the Nusselt number $Nu=qH/(\chi\Delta T \pi(D/2)^2)$, local temperatures, and the Reynolds number $Re=\frac{UH}{\nu}$ were measured as a function of $Ra$, $Pr$ and $\lambda$, where $q$ was the input heating power at the bottom plate, $\chi$ was the thermal conductivity of the working fluid and $U$ was the characteristic velocity of the LSC. Each measurement lasted at least 12 hours after the system had reached a stable state to ensure that the system had explored all flow states. 

\subsection{The viscous boundary layer measurements \label{SubSec_PIV}}

The dynamics of viscous boundary layer in the rough cell with $\lambda=1.0$ was measured using a commercial particle image velocimetry (PIV) system that had been documented elsewhere \citep*{Xia2003PRE}. The flow was seeded with particles with a diameter of 2 $\mu$m and density matched with water. A 2D flow region  with a size of 20 mm$\times$ 28 mm was measured  within the LSC's circulation plane with $x$ denoting the direction along the circulation path of the LSC and $z$ denoting the direction pointing upwards at the centre of bottom plate (see, e.g. figure \ref{fig_PIV}). The origin of the coordinate system was located at the centre of the plate. To minimise optical distortion of the particle images induced by the curved surface of the cell, a jacket filled with water with flat window was adhered to the convection cell. The optical arrangement was such that the measurement plane was within one groove of the rough plate and the images were acquired perpendicular to it. A mask with the shape exactly matching that of the plate geometry when viewing from the front was applied to the measured images to exclude the parts blocked by the roughness elements in the captured images.  

The velocity field was obtained by cross-correlating two consecutive images taken with a time separation that matches the velocity of the flow at different $Ra$. The velocity map consisted of 63$\times$ 79 vectors in the $x-$ and $z-$ directions respectively. Totally 26136 vector maps were acquired at a sampling rate of 2.2 Hz, which lasts for $\sim 3.3$ hours. The velocity fluctuation map was obtained by subtracting the mean velocity map from each of the instantaneous velocity map. The measurements were done at $Pr=4.34$ and $3.55\times10^8\leq Ra \leq 9.62\times10^9$. To lock the direction of the LSC, the convection cell was tilted with respect to the vertical direction by $\sim 0.5^o$. 

\section{Results and discussion \label{Sec_Results}}

\subsection{Heat transport measurements}\label{SubSec:Nu}

We show in this section that the heat transport in turbulent RBC over rough plates may be classified into three regimes based on the scaling law of $Nu$ vs. $Ra$. In addition, the heat transport scaling in the enhanced regime (Regimes II and III) can be manipulated by changing the roughness geometry. The transition from Regime I to Regime II in a rough cell with $h$ = 4 mm and $\lambda=4.0$ is discussed in section \S \ref{Regime_I_II}. The heat transport measurement with changing the geometry of roughness elements, i.e. for different values of $\lambda$, is presented in \S \ref{Regime_II_III} together with the observation of Regime II and Regime III. 

\subsubsection{Transition from Regime I to Regime II \label{Regime_I_II}}
\begin{figure}
\includegraphics[width=\textwidth]{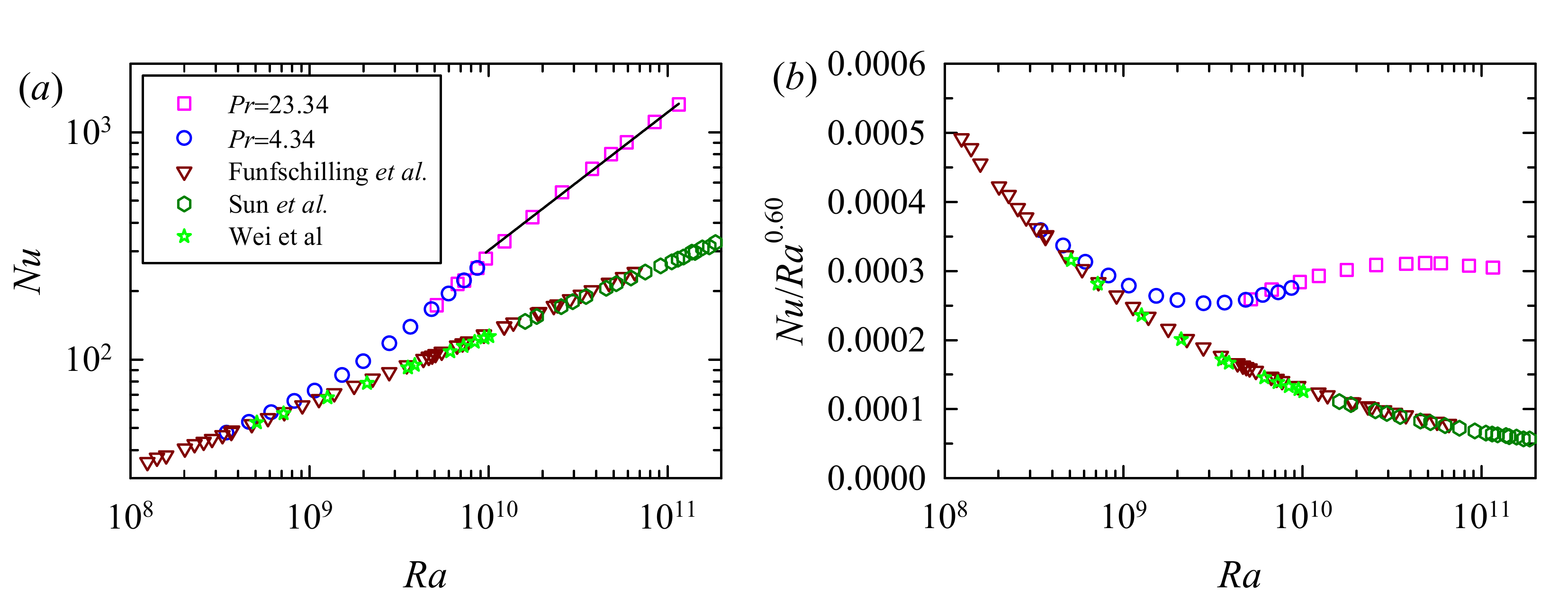}
\caption{\label{fig_Regime_I_II}(Colour online) ($a$) $Nu$ as a function of $Ra$ in a rough cell with a roughness elements height of 4 mm. The down-pointing triangles are taken from \citet{Denis2005JFM}($\Gamma=1.0, Pr=4.38$), the hexagons are taken  from \citet{Sun2005JFM}($\Gamma=2.0, Pr=4.34$), the stars are taken from \citet{Wei2014JFM}. ($b$) The compensated plot of the same data in ($a$) .}
\end{figure}

Previously, \citet{Shen1997PRL} find that the heat transport cannot be altered in a rough cell if thermal BL thickness $\delta_{th}$ is larger than the height of the roughness elements. We call this Regime I of convective turbulence in a rough cell. As $\delta_{th}$ decreases with increasing $Ra$, it will eventually become smaller than the roughness height, i.e. $\delta_{th} < h$. When this happens turbulent convection enters a new, heat transport enhanced regime, termed Regime II here. This regime has been observed by a number of previous experiments \citep{Shen1997PRL,Du1998PRL,Qiu2005JoT,Tisserand2011PoF,Wei2014JFM}.

We plot in figure \ref{fig_Regime_I_II}($a$) $Nu$ as a function of $Ra$ in a rough cells with $h=4$ mm and $\lambda=4.0$. For comparison, the data from \citet{Denis2005JFM} in a smooth cell with $\Gamma=1, Pr=4.38$, from \citet{Wei2014JFM} in a smooth cell with $\Gamma=1, Pr=4.34$ and from \citet{Sun2005JFM} in a smooth cell with $\Gamma=2, Pr=4.34$ are plotted respectively as down-pointing-triangles, hexagons and stars. Figure \ref{fig_Regime_I_II}($b$) plots the corresponding data in a compensated form. It is seen that $Nu$ in the rough cells first overlaps nicely with those from the smooth cells for $Ra < 4\times 10^8$. With increasing $Ra$, there is a gradual transition, where the heat transport enhances. Within the transitional $Ra$ range $ 4\times 10^8 < Ra < 1\times 10^{10}$, there is no well defined scaling law between $Nu$ and $Ra$. For $Ra>1\times 10^{10}$, the data shows a power-law relation, i.e. $Nu\sim Ra^{0.60}$. These data clearly demonstrate  the transition from Regime I to Regime II. Such a transition is also observed by \cite{Wei2014JFM} in a rough cell with $h=3$ mm.  Using the relation $\delta_{th}=H/(2Nu)$ (which was valid only for the plate-wise averaged thermal BL thickness and for a smooth cell), we had $\delta_{th}=2.1$ mm at $Ra=3.4\times 10^8$, which might be considered to be close to the roughness height of 4 mm in an order of magnitude sense and broadly consistent with the proposed interpretation by \citet{Shen1997PRL}.

\subsubsection{Effects of roughness geometry on the heat transport enhanced regime\label{Regime_II_III}}

We present measurements of $Nu$ as a function of $Ra$ in convection cells with a roughness height $h$ of 8 mm and different roughness geometries in this subsection. The roughness height was much larger than $\delta_{th}$ for the explored range of $Ra$. Therefore, Regime I behaviour was not observed here. Instead, we focus on Regime II and Regime III (a new regime to be defined below).

\begin{figure}
\includegraphics[width=\textwidth]{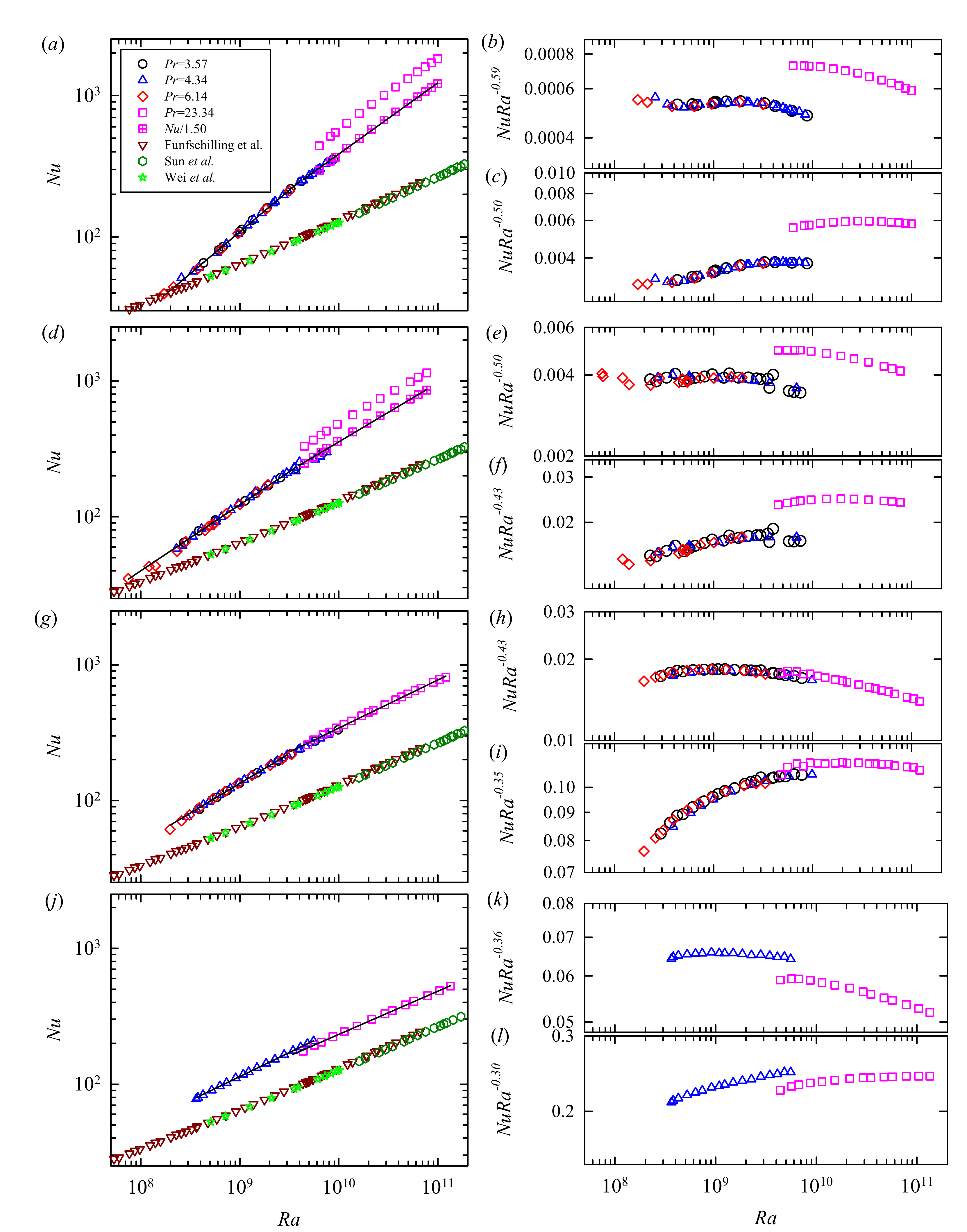}
\caption{\label{fig1}(Colour online) Left panel: $Nu$ as a function of $Ra$ in rough cells for ($a$) $\lambda=4.0$; ($d$) $\lambda=1.9$; ($g$) $\lambda=1.0$ and ($j$) $\lambda=0.5$. The down-pointing triangles are from \citet{Denis2005JFM}($\Gamma=1.0, Pr=4.38$), the hexagons are from \citet{Sun2005JFM}($\Gamma=2.0, Pr=4.34$) and the stars are from \cite{Wei2014JFM}($\Gamma=1.0, Pr=4.38$). Data at $Pr=4.34$ in ($j)$ are from \citet{Wei2014JFM}; $Nu$ at $Pr=23.34$ at $\lambda = 4.0$ and 1.9 are larger than those at $Pr = 3.57$ at the same $Ra$. For a better comparison, they are shifted downwards respectively by 1.5 and 1.34 and displayed using squares with cross. The solid lines are power law fits to the data larger and smaller than $Ra^*\sim 4\times 10^9$. Right panel: The compensated plot of $Nu$ with the fitted $Ra$ scaling for $Ra$ larger and smaller than $Ra\sim 4\times 10^9$  from the left panel.}
\end{figure}

Figures \ref{fig1}($a, d, g, j$) show $Nu$ as a function of $Ra$ and $Pr$ in rough cells with $\lambda=4.0, 1.9, 1.0$ and 0.5, respectively. The solid lines in the figure are power law fit to each data set for $Ra$ larger and smaller than a critical Rayleigh number $Ra^*=4\times 10^9$. The same data from smooth cells as shown in figure \ref{fig_Regime_I_II} are shown in these figures. The general conclusion that one can draw from these figures is that the heat  transport is enhanced in rough cells. 

We first discuss the case with $\lambda=1.0$ (figure \ref{fig1}($g$)). Within the precision of the experiments, data at different values of $Pr$ but the same $Ra$ collapse onto each other, suggesting that the $Pr$-dependence in the rough cell with $\lambda=1.0$ is very weak in the present parameter range, which is consistent with observation in a smooth cell \citep*{Xia2002PRL}. Detailed examination of the data shows that there is a transition of the $Nu-Ra$ scaling at $Ra^*\sim 4\times 10^{9}$. Below and above $Ra^*$, the data can be fitted by different scaling laws, i.e. $Nu\sim Ra^{0.43\pm 0.01}$ and $Nu\sim Ra^{0.35\pm 0.01}$. The different scaling laws suggest that the heat transport enhancement regime in a rough cell can be further classified into two regimes for the $Pr$ range covered in the present experiments. We term the two regimes before and after $Ra^*$ as Regime II and Regime III. The data are compensated by the scaling exponents in Regimes II and III and shown in figure \ref{fig1}($h,i$). A gradual transition around $Ra^*\sim 4\times 10^9$ is clearly seen.

Data at $\lambda=4.0, 1.9$ and 0.5 are plotted in figure \ref{fig1} ($a, d, j$). One interesting feature revealed in figure \ref{fig1}($a$) and ($d$) is that the $Nu$ at $Pr=23.34$ is larger than that at $Pr=3.57$ when $Ra$ is the same. We note that this $Nu$-enhancement for FC770 ($Pr=23.34$), as compared to water ($Pr=3.57$), only occurs for the large $\lambda$ cases. In fact, for the smallest value of $\lambda$ ($=0.5$) there is a slight decrease in $Nu$. This is quite surprising and unexpected, as in the case of a smooth cells, $Nu$ is peaked around $Pr=3 \sim 4$ \citep{Xia2002PRL}. It is not clear why the heat transport enhancement in rough cells has this peculiar $Pr$-dependence, and furthermore this behaviour has a strong $\lambda$ dependence, i.e. its 150\% for $\lambda=4.0$, 134\% for $\lambda=1.9$, 4\% for $\lambda=1.0$, and -8\% for $\lambda=0.5$. One possible reason leading to such a $Pr$-abnormality is that the lifetime of the thermal plumes at larger $Pr$ is larger compared to the small $Pr$ case. Assuming that the thermal plumes have typical length scale comparable to the thermal BL thickness, the estimated thermal diffusion time scale, i.e. $\delta_{th}^2/\kappa$, for FC770 is 14.5 sec and that for water is 2.6 sec at $Ra\approx 10^{10}$. When normalised by the LSC turn-over time, the dimensionless plume lifetime for FC770 is 0.25 and that for water is 0.11 for $\lambda=4.0$. Another likely reason is the stronger plume clustering effect in rough cells with larger $\lambda$ as shown by the shadowgraph images in figure \ref{fig_shadow}, which also has the effect of increasing plume lifetime or slowing the decay of plumes. While it is difficult to quantify the combined effect of these two factors, it appears that larger values of $\lambda$ together with a larger $Pr$ result in a more efficient enhancement in the heat transport. For small $\lambda$ and/or small $Pr$, thermal plumes lose more heat to the surrounding fluid and therefore $Nu$ won't be enhanced as much.

\begin{figure}
\begin{center}
\includegraphics[width=0.65\textwidth]{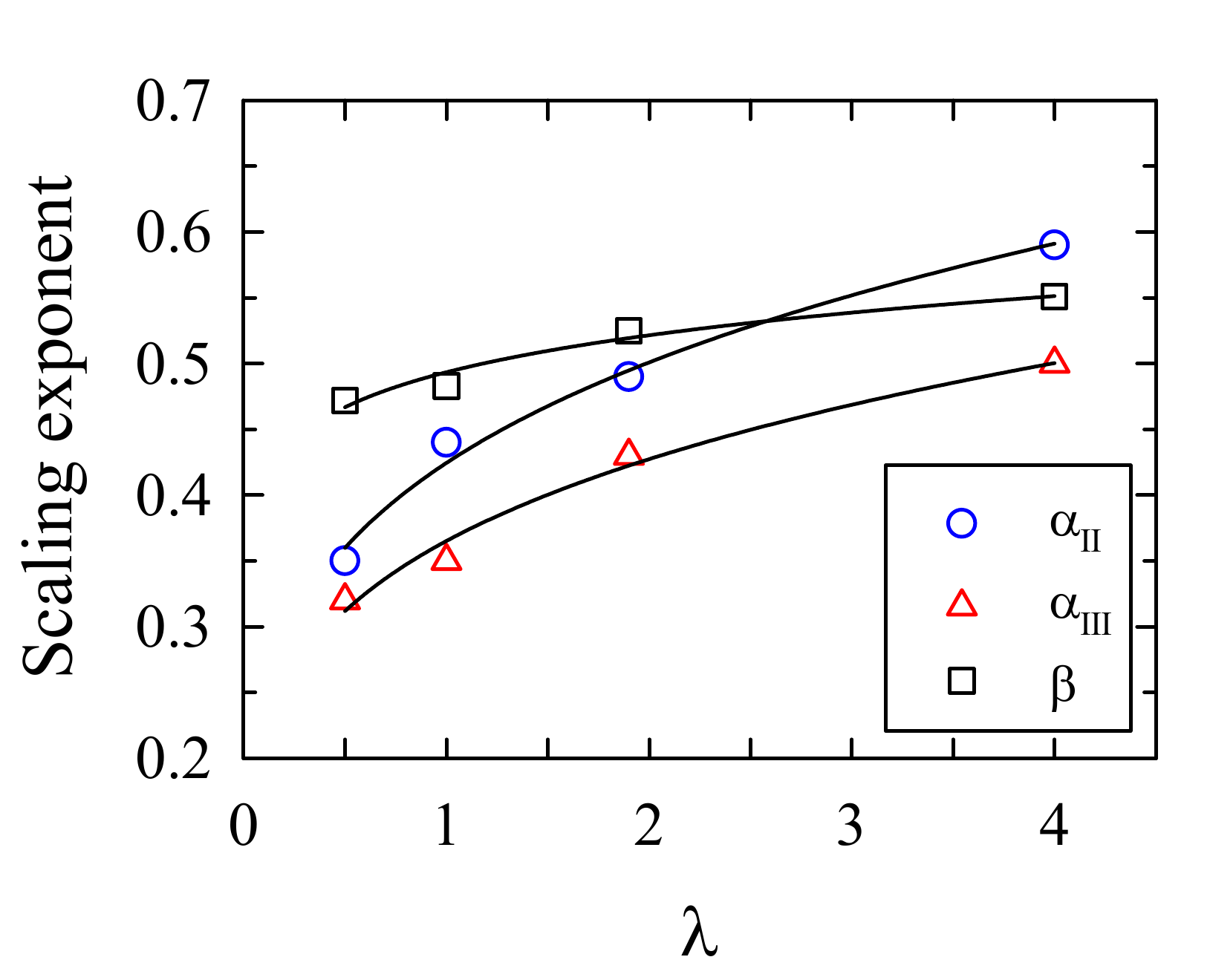}
\caption{\label{fig3}(Colour online) The scaling exponent $\alpha$ from power law fits to $Nu$ vs $Ra$, i.e. $Nu\sim Ra^{\alpha}$, in Regime II (circles) and Regime III (triangles) as a function of $\lambda$. Also plotted in the figure is the scaling exponent $\beta$ from power law fits to $Re$ vs $Ra$ (squares), i.e. $Re\sim Ra^{\beta}$, as a function of $\lambda$. The solids lines are empirical power law fits to the respective data sets: $\alpha_{II}=0.43\lambda^{0.24\pm 0.02}$, $\alpha_{III}=0.37\lambda^{0.23\pm 0.02}$ and $\beta=0.49\lambda^{0.08\pm0.01}$.}
\end{center}
\end{figure}

To determine more accurately the scaling law of $Nu$ vs. $Ra$, the data at $Pr=23.34$ in figure \ref{fig1}($a$) and ($d$) are shifted downwards by a constant, i.e. 1.50 in ($a$) and 1.34 in ($d$), to collapse data at different $Pr$ but the same $Ra$. The transition of the heat transport scaling around $Ra^*$ is observed for all $\lambda$ studied. The power law fits of $Nu=ARa^{\alpha}$ in Regimes II and III for different values of $\lambda$ are summarised in table \ref{tab:table1}. It should be noted that the shifted high $Pr$ data were included in the fitting. Since $Ra$ obtained using water as working fluid at $Pr=3.57$ and that using FC770 as working fluid at $Pr=23.34$ had an overlapping range of less than a half decade, i.e. from $6\times 10^9$ to $9\times 10^9$, a determination of the $Pr$ dependence in Regime III was not possible in the present experiments. It is seen that the heat transport law is distinctively different in Regime II and Regime III. This can be seem more clearly from the compensated plot in the right panel next to figure \ref{fig1}($a,d,j)$. 

The experimentally determined scaling exponent $\alpha$, i.e. $Nu\sim Ra^{\alpha}$, in both Regime II and Regime III increases significantly with $\lambda$. In Regime II, the $\alpha$ increases from 0.36 to 0.59 when $\lambda$ is increased from 0.5 to 4.0; and in Regime III, it increases from 0.30 to 0.50. These data are plotted in figure \ref{fig3}. If a power law fit is attempted to these data within the limited range of $\lambda$, we obtain $\alpha_{II}=0.42\lambda^{0.23\pm 0.02}$ and $\alpha_{III}=0.35\lambda^{0.24\pm 0.02}$ which are shown as solid lines in the figure. Within the error bar, the $\lambda$-dependences of $\alpha_{II}$ and $\alpha_{III}$ are the same except that the pre-factor of $\alpha_{II}$ is larger than $\alpha_{III}$. The results suggest that the scaling law of the heat transport in turbulent RBC over rough plates can be manipulated using $\lambda$. In this sense, the present experiments demonstrate clearly the importance of roughness geometry on turbulent heat transport. Any realistic model for turbulent thermal convection in nature should take into account the effects of wall roughness. 

Very recently, it is found that there is an upper bound for the heat transport in turbulent convection over rough plates, i.e. $Nu$ cannot grow faster than $Nu\sim Ra^{1/2}$ in the limit of $Ra\rightarrow\infty$ \citep{Goluskin2016JFM}. With increasing $Ra$, Regime II is just a transient, and Regime III is the dominated one. Our experiments showed that at $\lambda=4.0$ the scaling exponent already reached 0.50, which was consistent with the upper bound derived \citep{Goluskin2016JFM}. One interesting question is whether the scaling exponent will saturate to 0.50 with further increasing $\lambda$ which is very challenging to achieve experimentally due to technical difficulties in machining the plates. DNS studies may provide helpful insights into this aspect of the problem.

\begin{figure}
\begin{center}
\includegraphics[width=0.65\textwidth]{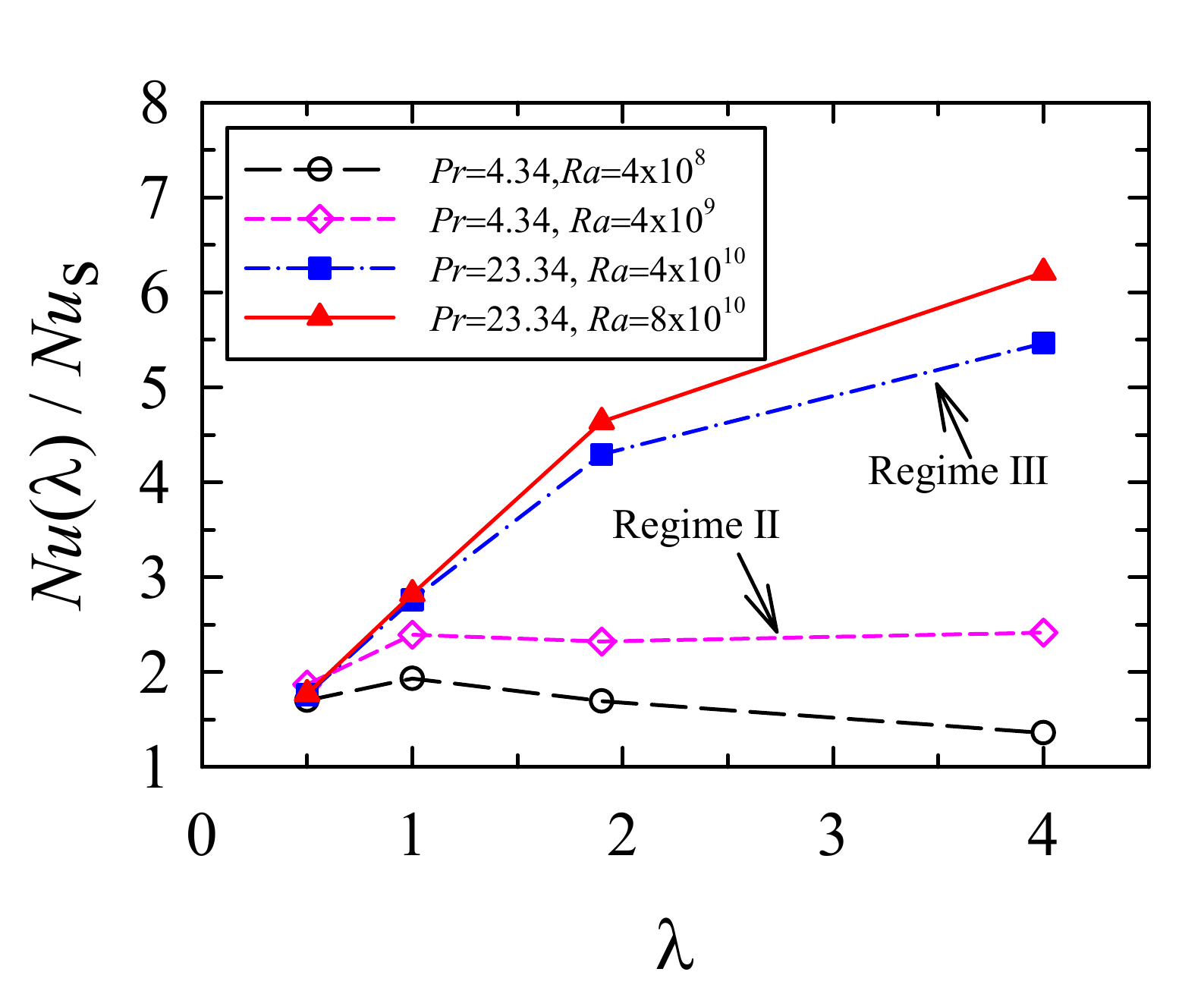}
\caption{\label{fig_enhancement}(Colour online) The heat transport enhancements in Regime II (open symbols) and Regime III (solid symbols) as a function of $\lambda$.}
\end{center}
\end{figure}

\begin{table}
  \begin{center}
\def~{\hphantom{0}}

\begin{tabular}{cccc}
 \textrm{$\lambda$}&\tabincell{c}{$Nu=ARa^{\alpha}$\\Regime II \hspace{0.5 cm} Regime III}&$Re=BRa^{\beta}Pr^{\epsilon}$&\tabincell{c}{$\sigma_T/\Delta T=C\times Ra^{\gamma}$\\center \hspace{0.5 cm} sidewall} \\
 
   4.0 & $0.00052Ra^{0.59}$\quad$ 0.004Ra^{0.50 }$ &$0.088Ra^{0.551}Pr^{-0.66}$&  $0.08Ra^{-0.09}$\quad$0.138Ra^{-0.08}$   \\  
    1.9 & $0.005Ra^{0.50}$\qquad$0.016Ra^{0.43}$   &$0.169Ra^{0.525}Pr^{-0.71}$& $0.19Ra^{-0.13}$\quad$0.135Ra^{-0.09}$  \\
1.0 &$0.015Ra^{0.43}$ \quad$0.10Ra^{0.35}$        &$0.395Ra^{0.486}Pr^{-0.69}$& $0.49Ra^{-0.18}$\quad$0.92Ra^{-0.18}$   \\
   0.5 & $0.74Ra^{0.36}$ \qquad$0.15Ra^{0.30}$   &$0.481Ra^{0.472}Pr^{-0.72}$&$0.24Ra^{-0.16}$\quad$3.60Ra^{-0.23}$  \\  
\end{tabular}
 \caption{\label{tab:table1} Measured scaling law of the Nusselt number $Nu$ and normalized temperature fluctuations $\sigma_T/\Delta T$ in Regime III for different values of $\lambda$.}
  \label{tab:kd}
  \end{center}
\end{table}

We next study the $\lambda$-dependence of the heat transport enhancement behaviour in Regime II and Regime III. Figure \ref{fig_enhancement} shows the heat transport enhancement in rough cells compared with that in a smooth cell. It is seen that the heat transport enhancement in Regime II (open symbols in figure \ref{fig_enhancement}) first increases and then decreases with increasing $\lambda$. The situation is very different in Regime III (solid symbols in figure \ref{fig_enhancement}) where it is found that the heat transport enhancement increases monotonically with increasing $\lambda$. A maximum heat transport enhancement of 637\% is observed at $\lambda=4.0$. 

We note that in a DNS study, \citet{Wanger2015JFM} show that the heat transport is enhanced more for slender roughness elements, i.e. larger $\lambda$, using only four roughness elements with a roughness height of 0.125H (``obstales" in \cite{Wanger2015JFM}) in a quasi-2-D convection cell. The observation in \cite{Wanger2015JFM} is qualitatively in agreement with what we find here, e.g. heat transport enhancement generally increases with $\lambda$ (see e.g. figure \ref{fig_enhancement}). 

\subsection{Transition from Regime II to Regime III: The viscous boundary layer cross-over \label{SubSec_Transition}}

To understand what may have happened at $Ra^*$, i.e. the transition from Regime II to Regime III, we recall that there are two BLs in turbulent RBC, i.e. the thermal BL and the viscous BL. The viscous BL is thicker than the thermal BL for $Pr>1$. According to the Grossmann-Lohse model for turbulent RBC, BLs play crucial role in determine the global heat transport \citep{GL2000JFM}. The transition seen in figure \ref{fig1} may be understood in terms of the viscous BL cross-over, i.e. the roughness elements not only strongly perturb the thermal BL, but also alter the viscous BL in Regime III. To verify this, we directly measured the dynamics of the viscous boundary layer in a rough cell with $\lambda=1.0$.

\begin{figure}
\begin{center}
\includegraphics[width=\textwidth]{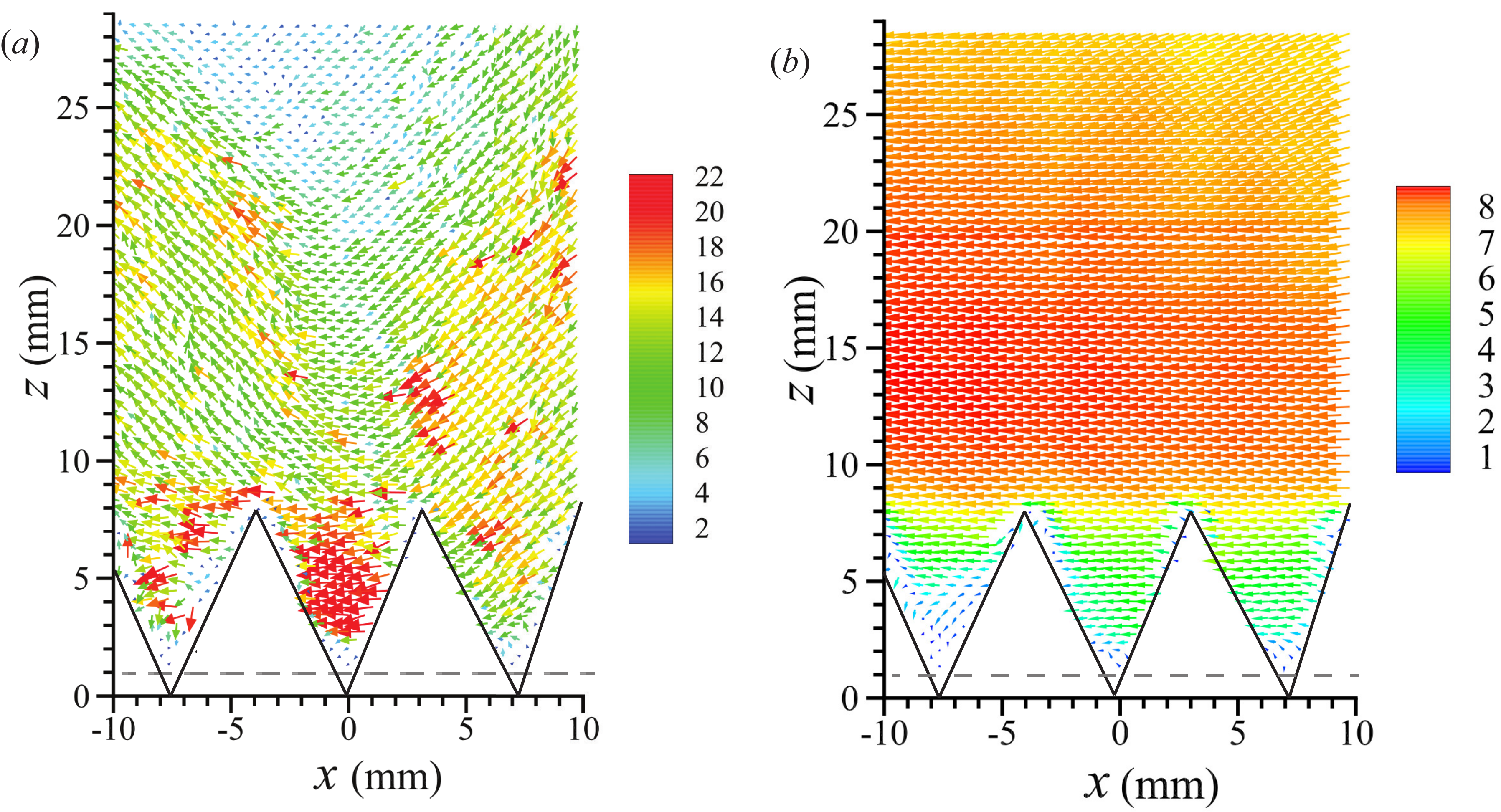}
\caption{\label{fig_PIV}(Colour online) The instantaneous ($a$) and time-averaged ($b$) velocity fields measured near the bottom plate within the large-scale circulation plane ($Ra=4.22\times 10^9$, $Pr=4.34$ and $\lambda=1.0$). The velocity magnitude (in units of mm/s) is colour coded with the colour bar nearby each figure, and also represented by the length of the vectors. The velocity field is not accessible  experimentally below the dashed lines.}
\end{center}
\end{figure}

\begin{figure}
 \centering
\includegraphics[width=\textwidth]{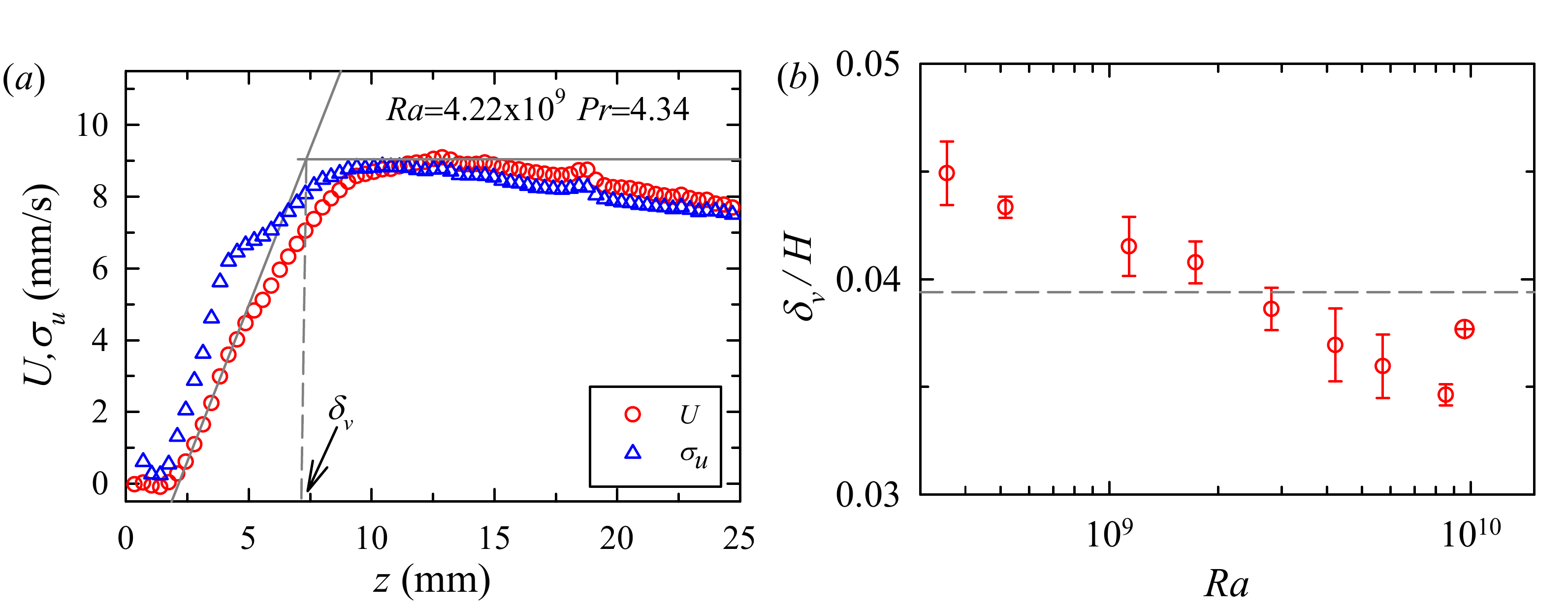}
\caption{\label{fig2} (Colour online)($a$) Typical time-averaged horizontal velocity $U$ and root-mean-square horizontal velocity $\sigma_u$ profiles ($Ra= 4.22\times10^9$, $\lambda=1.0$ and $Pr=4.34$). The vertical dashed line indicates the location of the viscous boundary layer. ($b$) Normalized viscous boundary layer thickness $\delta_v/H$ vs $Ra$. The dashed line marks the height of the roughness elements. The datum at the highest $Ra$ (symbols with cross, $\Delta T= 29.83K$) shows different behaviours with other data points which we believe is an artefact due to the strong optical distortion of the images used for PIV analysis.}
\end{figure}

\begin{figure}
\centering
\includegraphics[width=\textwidth]{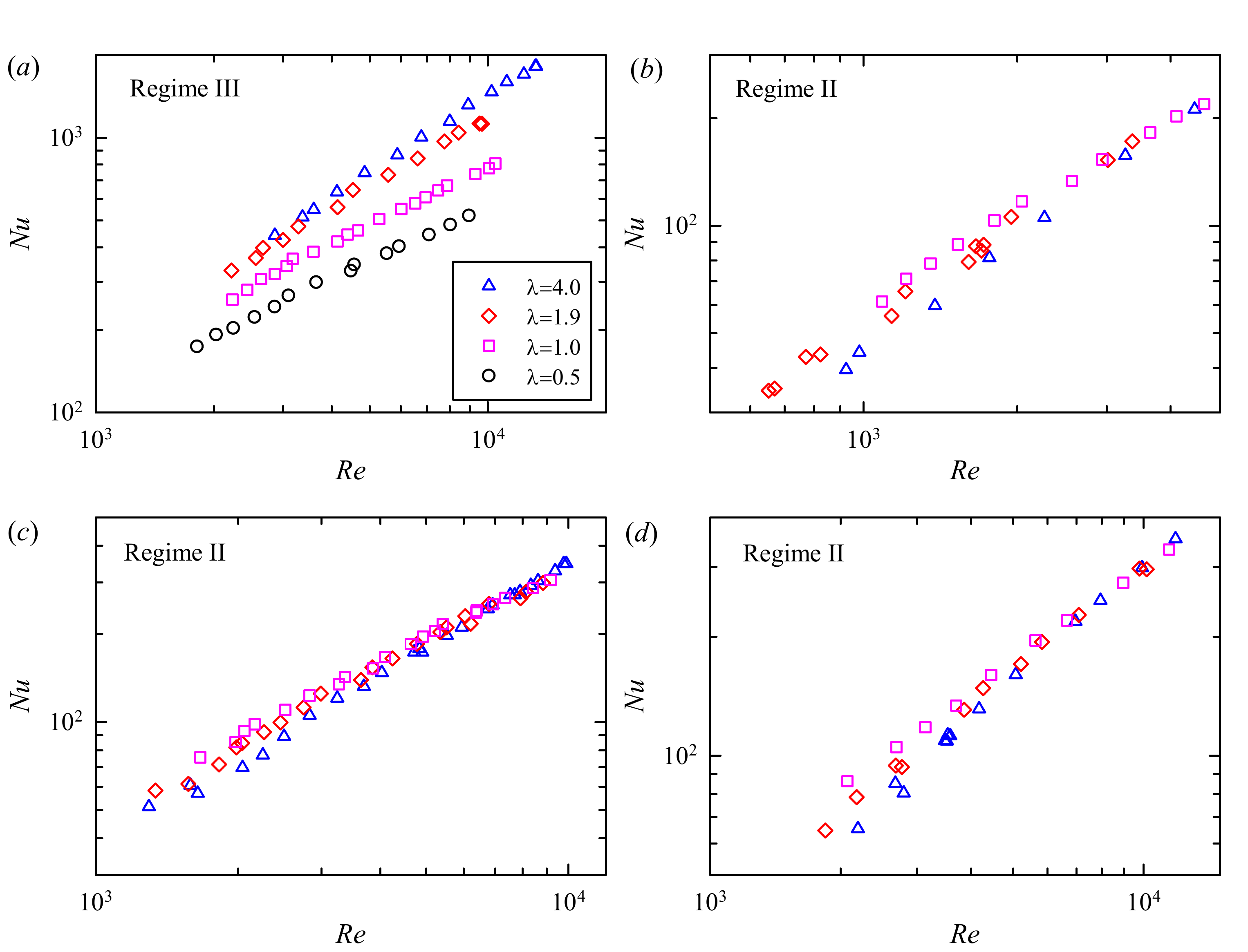}
\caption{\label{fig_Nu_Re} (Colour online) $Nu$ as a function of $Re$ measured in rough cells with different values of $\lambda$ at ($a$) $Pr=23.34$ ($b$) $Pr=6.14$ ($c$) $Pr=4.34$ and ($d$) $Pr=3.57$.}
\end{figure}

An example of the instantaneous velocity map measured at $Ra=4.22\times10^{9}$ and $Pr=4.34$ is shown in figure \ref{fig_PIV}($a$). The triangles outline the space occupied by the roughness elements schematically. The velocity magnitude is coded by both the colour and the length of the vectors. The instant velocity field shows that the mean flow is going from the right to the left and there are intermittent bursts of the flow velocity. Figure \ref{fig_PIV}($b$) depicts the time averaged velocity field with the velocity magnitude coded also in both colour and the length of the vectors. One may notice that the time-averaged maximum velocity magnitude is significantly reduced when compared with the instantaneous one as revealed by the scale bars. After averaging over time one sees that the mean velocity is uniform with respect to $x$ except very close to the plate, e.g. in between the roughness elements. A tiny counter-rotating vortex may be identified in the region -8 mm $< x < $ -5 mm and 0 mm $ < z < $ 4 mm. The mean velocity field varied weakly with $x$ (figure \ref{fig_PIV}($b$)). We thus averaged the mean velocity field along $x-$direction to obtain the mean velocity and the root-mean-square (rms) velocity profiles.

Figure \ref{fig2}($a$) shows an example of the horizontal mean velocity profile $U(z)$ and the profile of the rms velocity $\sigma_U(z)$ measured at $Ra=4.22\times 10^9$ and $Pr=4.34$. A remarkable feature shown in figure \ref{fig2}($a$) is that the rms velocity is comparable to the mean velocity very close to the plate, and it's even larger than the mean velocity within the viscous BL. The origin of the enhanced velocity fluctuation is not clear but speculatively this should largely correlate with the intermittent bursts of thermal plumes. In contrast, the maximum rms velocity in a smooth cell is only about 30\% of the maximum mean velocity and it is always smaller than the mean velocity within the boundary layer region \citep*{Sun2008JFM, Wei2013JFM}. 

The viscous boundary layer thickness was determined using the so-called slope method: The intersection between the extrapolation of the linear part of the velocity profile very close to the plate and the maximum mean horizontal velocity was defined as the thickness of the viscous boundary layer $\delta_v$ (indicated by the vertical dashed line in figure \ref{fig2}($a$)) \citep{XinXiaTong1996PRL,XinXia1997PRE,QiuXia1998PRE,Sun2008JFM, Wei2013JFM}. The $Ra$-dependence of the viscous BL thickness normalised by the height of the convection cell is plotted in figure \ref{fig2}($b$). With increasing $Ra$, $\delta_u/H$ becomes thinner and thinner. At $Ra \sim 3\times 10^9$, $\delta_u/H$ approximately equals to $h/H$. This roughly corresponds to $Ra^*$ seen in the measured $Nu$ ($Ra^*\sim 4\times 10^9)$. This observation indicates that the transition of the heat transport law between Regimes II and III may be caused by the viscous BL becoming smaller than the height of the roughness elements. After this viscous BL cross-over, the roughness elements strongly perturb both the thermal and the viscous BLs. The thermal plumes emitted from the tip of pyramids can then be directly ejected into the bulk. For $Pr=23.34$, the estimated viscous BL thickness at the smallest $Ra$ explored in the experiments ($Ra=4.39\times 10^9$) was 0.04H \citep{Lam2002PRE}, which was very close to the height of roughness element, implying that the measurements with $Pr=23.34$ were all in Regime III. 

In the asymptotically large $Ra$ limit, the thickness of the BLs is so small such that the roughness elements perturb both the thermal and the viscous BLs. The system is in Regime III in this case wherein $Nu$ is very sensitive to the LSC, or equivalently to the Reynolds number $Re$ associated with the LSC (to be defined in section \ref{SubSec_Re}). We plot in figure \ref{fig_Nu_Re} $Nu$ as a function of $Re$ for ($a$) $Pr$ = 23.34, ($b$) $Pr$ = 6.14, ($c$) $Pr$ = 4.34, and ($d$) $Pr$ = 3.57. On one hand, figure \ref{fig_Nu_Re} ($a$) shows that $Nu$ increases with $\lambda$ for all $Re$ measured at $Pr=23.34$, which are in the Regime III. On the other hand, $Nu$ at other $Pr$ where mostly within the Regime II shows weak dependence on  $Re$. This is another evidence supporting that the viscous BL cross-over may also play important role in turbulent thermal convection with the presence of roughness elements.

\subsection{Reynolds number measurements\label{SubSec_Re}}

\begin{figure}
\centering
\includegraphics[width=0.65\textwidth]{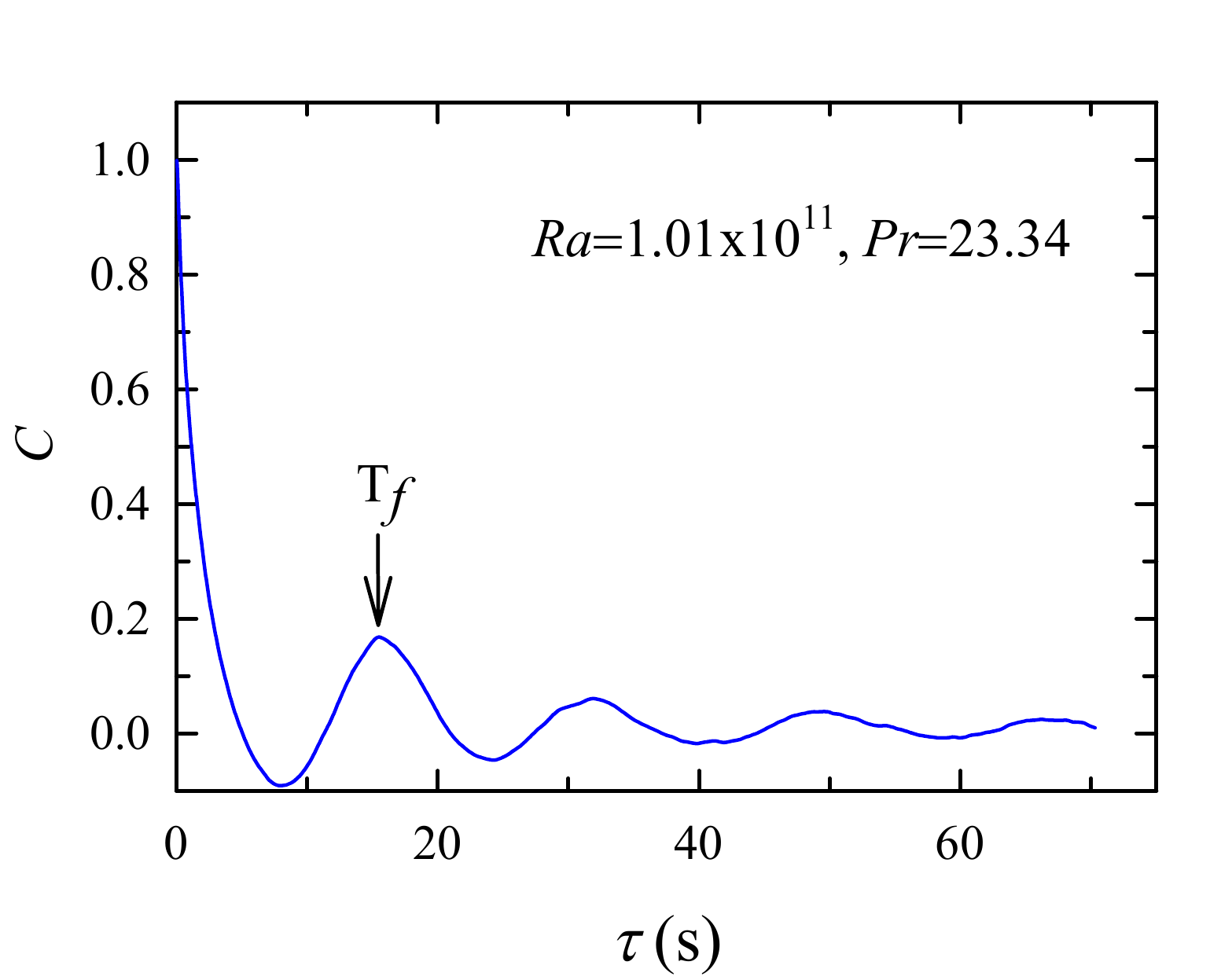}
\caption{\label{fig_Oscillation} Autocorrelation function of the temperature measured $\sim 1$ cm away from sidewall in a rough cell with $\lambda=4.0$ at $Ra=1.01\times 10^{11}$ and $Pr=23.34$. Here $T_f$ is the turnover time of the LSC.}
\end{figure}

The self-organised large-scale circulation (LSC, or ``wind of turbulence'') is one of the fascinating features of turbulent RBC. The flow dynamics of the LSC is characterised by the associated Reynolds number $Re$. We study in this section how the roughness geometry alters the dynamics of the LSC by examining the $Re$ dependence on both $Ra$ and $Pr$. It has been well-known that there exits a well-defined low-frequency oscillation in turbulent RBC. The time scale of this oscillation is interpreted as the turnover time $T_f$ of the LSC (see, e.g. \citet*{Xie2013JFM} and references therein). By choosing a typical length scale of the LSC, e.g. 4H, the Reynolds number is calculated as:
\begin{equation}
Re=UH/\nu=4H^2/(T_f\nu)
\end{equation}
Time-resolved temperature signals measured $\sim 1$cm away from the sidewall were used to determine $T_f$ and thus the $Re$ as a function of $Ra$ and $Pr$ when changing the roughness geometry, i.e. at different values of $\lambda$. The turnover time of the LSC was determined by locating the time position of the first peak of the temperature auto-correlation function $C(\tau)$, which is defined as:
\begin{equation}
C(\tau)=\frac{\langle[T(t+\tau)-\langle T(t) \rangle_t ][T(t)-\langle T(t)\rangle_t]\rangle_t}{\sigma_T^2}
\end{equation}
where $\langle\cdots\rangle_t$ denoted time-averaging and $\sigma_T$ is the rms temperature. An example of the autocorrelation function measured in convection cell with $\lambda=4.0$ at $Ra=1.01\times 10^{11}$ and $Pr=23.34$ is depicted in figure \ref{fig_Oscillation}. Clear oscillations can be seen from $C(\tau)$. Fitting the data points nearby the first peak of $C(\tau)$ with a six order polynomial, the time location of the first peak was determined and denoted as $T_f$. The uncertainty of $T_f$ was from $0.06$ s to $0.008$ s with increasing $Ra$.

\begin{figure} 
\centering
\includegraphics[width=\textwidth]{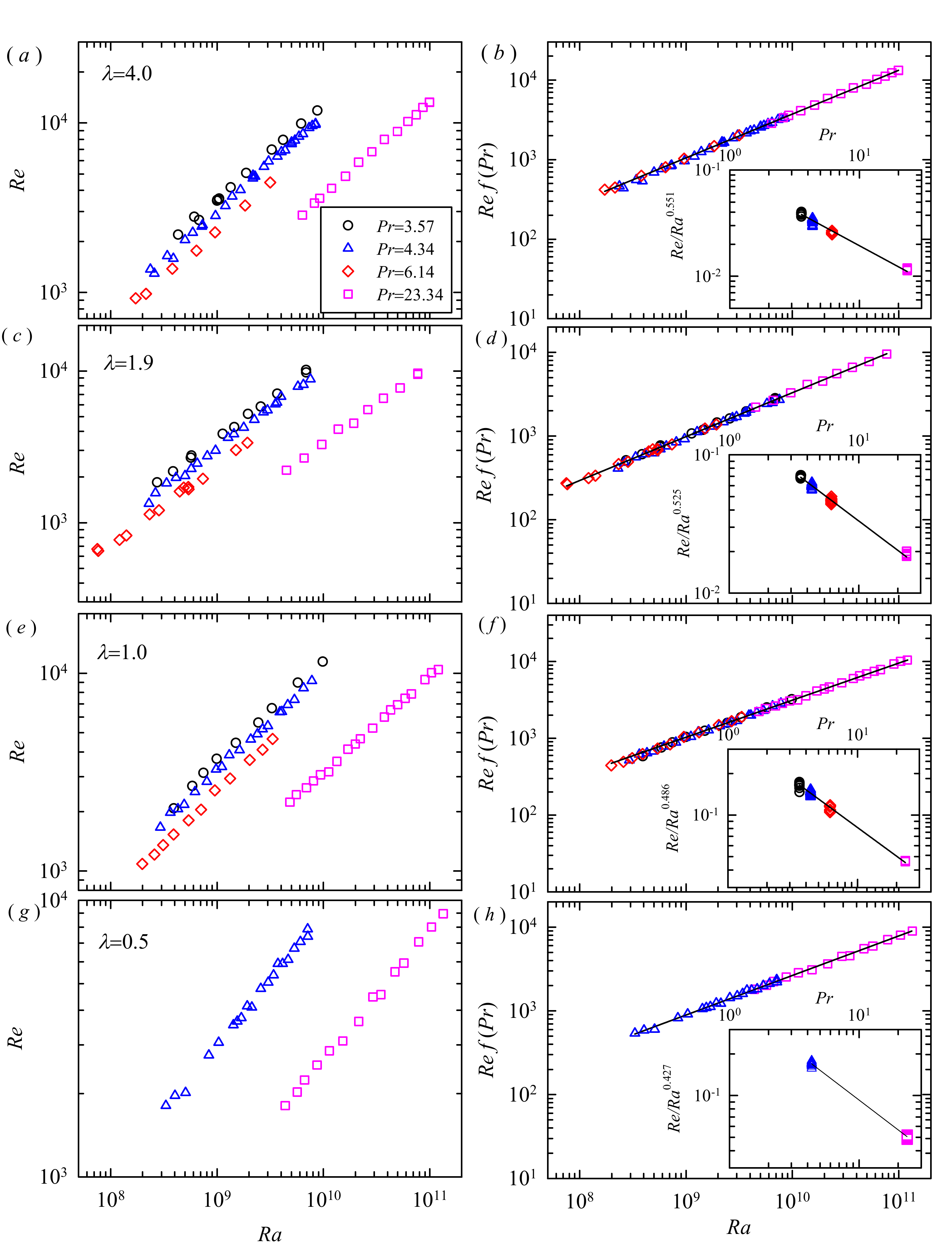}
\caption{\label{fig_Re} (Colour online) Reynolds number $Re$ as a function of $Ra$ and $Pr$ in rough cells for ($a$) $\lambda=4.0$, ($c$) $\lambda=1.9$, ($e$) $\lambda=1.0$ and ($g$) $\lambda=0.5$. To determine more precisely the scaling exponent of $Re$ with respect to $Ra$, data measured at $Pr=3.53, 4.34$ and $6.14$ in ($a, c, e, g$) are multiplied by a $Pr$-dependent constant $f(Pr)$ to collapse with data at $Pr=23.34$. The shifted data are shown in the main plot in the right panel ($b, d, f, g$) with power law fit $Re f(Pr)\sim Ra^{\beta}$ shown as solid lines. The insets in ($b, d, f, g$) show the compensated plot $ReRa^{-\beta}$ with respect to $Pr$ with the solid lines representing power law fits to the data. See table \ref{tab:table1} for the fitted power laws.}
\end{figure}

Figure \ref{fig_Re} displays $Re$ as a function of $Ra$ and $Pr$ in rough cells with $\lambda=4.0$  ($a,b$), $\lambda=1.9$ ($c,d$),  $\lambda=1.0$ ($e,f$) and $ \lambda=0.5$ $(g,h)$. Let's first discuss the case for $\lambda=4.0$. The $Re$ as a function of $Ra$ for $Pr=3.57, 4.34, 6.14$ and $23.34$ are plotted in figure \ref{fig_Re}($a$) on a log-log scale. The $Re$ exhibits power law dependence on $Ra$ at different $Pr$, i.e. $Re\sim Ra^{\beta}$. To determine more precisely $\beta$, data at $Pr$ = 3.57, 4.34 and 6.14 were multiplied by a $Pr$-dependent constant to collapse them with data at $Pr=23.34$ onto a single line. The shifted data are shown in the main plot of figure \ref{fig_Re}{($b$)}. A power law fits to the data over almost three decades of $Ra$ yielded: $Re\sim Ra^{0.551\pm0.001}$. The $Ra$-dependence of $Re$ was then divided out by compensating the data in figure \ref{fig_Re}($a$) with the so-determined power law, e.g. the inset of figure \ref{fig_Re}($b$). Within a relatively limited range of $Pr$, the data was also described by a power law, i.e. $Re/Ra^{0.551}\sim Pr^{-0.66\pm 0.02}$. The so-determined $Re(Ra, Pr)$ was:

\begin{equation}
Re=0.088Ra^{0.551}Pr^{-0.66}\quad (\lambda=4.0)
\end{equation}

Applying the same procedure to data obtained in rough cells with $\lambda=1.9$ (figure \ref{fig_Re}($c$)), $\lambda=1.0$ (figure \ref{fig_Re}($e$)), and $\lambda=0.5$ (figure \ref{fig_Re}($g$)), we determined the $Ra$ and  $Pr$ dependence of $Re$ for each values of $\lambda$,  respectively. The scaling laws of $Re=BRe^{\beta}Pr^{\epsilon}$ for different values of $\lambda$ are summarised in table \ref{tab:table1}. 

On one hand, with $\lambda$ increasing from 0.5 to 4.0 the scaling exponent $\beta$, i.e. $Re\sim Ra^{\beta}$, increases from 0.472 to 0.551, which is plotted as a function of $\lambda$ in figure \ref{fig3}. The data may be described by an empirical power law within very limited range of $\lambda$, i.e. $\beta=0.49\lambda^{0.08\pm0.01}$. On the other hand, the scaling exponent $\epsilon$, i.e. $Re\sim Pr^{\epsilon}$, remains at the same value around $-0.70$ within experimental uncertainty except for $\lambda=4.0$. Whether there is a change of the $Pr$-dependence of $Re$ in the rough cell with $\lambda=4.0$ requires future investigation with a wider range of $Pr$. The change of $\beta$ with $\lambda$ suggests a change of the LSC dynamics when the geometry of the roughness elements is changed. It is thus interesting to study torsional and sloshing oscillations of the LSC dynamics in rough cells with different values of $\lambda$ as those done in a smooth cell  \citep{Funfschilling2004PRL,Xi2009PRL,Xie2013JFM}. This is beyond the scope of the present studies and will be investigated in the future. Interestingly, the transition of the heat transport scaling law from Regime II to Regime III is not observed in the Reynolds number measurement.

\begin{figure}
\centering
\includegraphics[width=\textwidth]{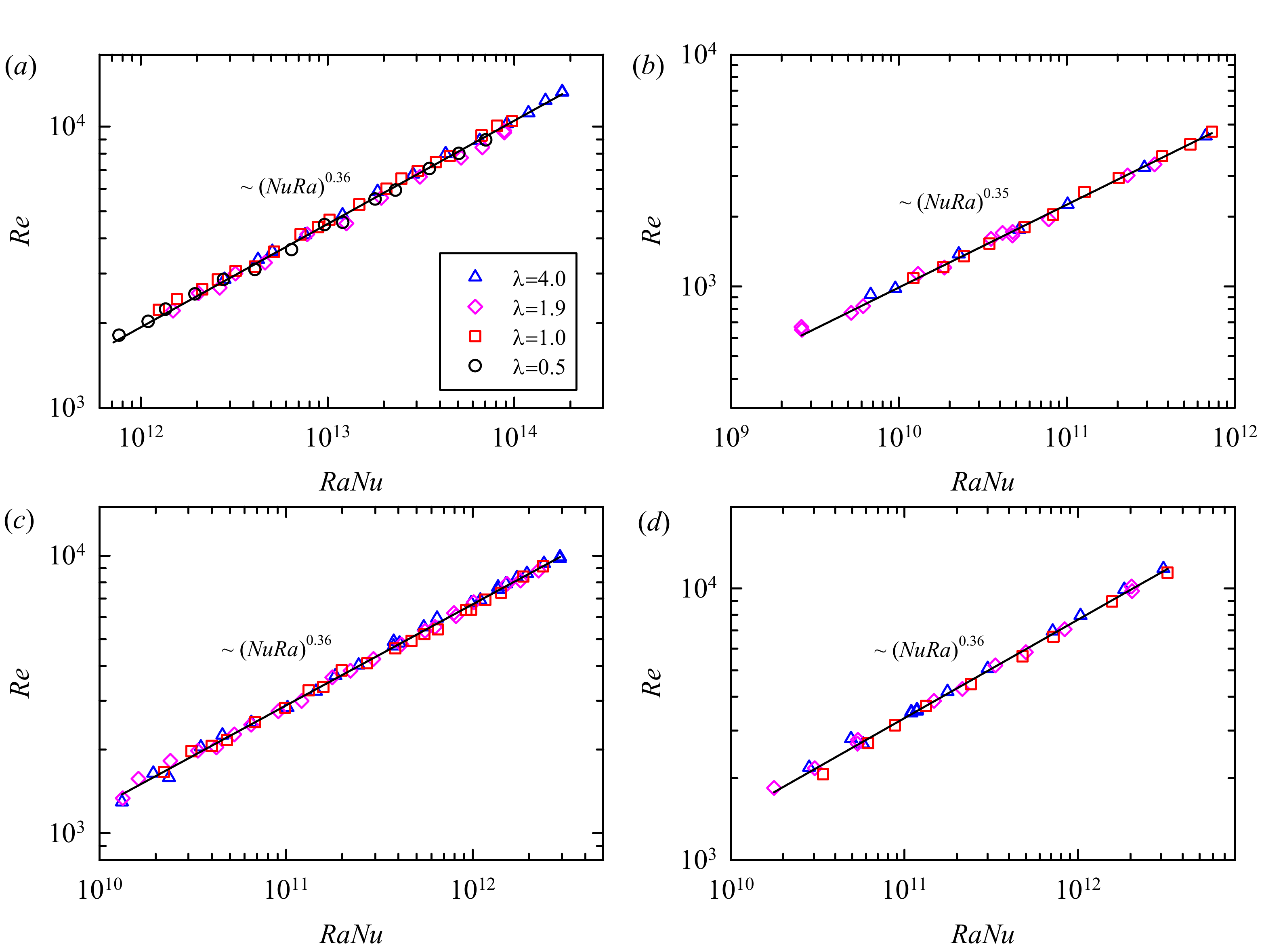}
\caption{\label{fig_ReRaNu} (Colour online) Reynolds number $Re$ as a function of $RaNu$ for different values of $\lambda$ with the solid line representing a power law fit to the data irrespective of $\lambda$. ($a$) $Pr=23.34$ and $Re=0.076\times (RaNu)^{0.36\pm0.01}$; ($b$) $Pr= 6.14$ and $0.266\times (RaNu)^{0.35\pm0.01}$; ($c$) $Pr= 4.34$ and $0.281\times (RaNu)^{0.36\pm0.01}$; ($d$) $Pr=3.57$ and Re=$0.340\times (RaNu)^{0.36\pm0.01}$.}
\end{figure}

It is found very recently that $Re$ is insensitive to the plate morphology for a given input heat flux for $\lambda=0$ and 0.5 \citep{Wei2014JFM}, i.e. when $Re$ is plotted against the flux Rayleigh number $RaNu$, data at different values of $\lambda$ collapse. We plot $Re$ as a function of $RaNu$ for different values of $\lambda$ and $Pr$ in figure \ref{fig_ReRaNu}. The data at the same $Pr$ but different $\lambda$ indeed overlap onto each other which is in agreement with the observation in \citet{Wei2014JFM}, but now for a wider range of $\lambda$ and $Pr$. In addition, data at the same $Pr$ can be fitted by a power law, i.e. $Re=0.076\times (RaNu)^{0.36\pm0.01}$ $(Pr=23.34)$, $Re=0.266\times (RaNu)^{0.35\pm0.01}$ $(Pr = 6.14)$, $Re=0.281\times (RaNu)^{0.36\pm0.01} (Pr=4.34) $, and $Re=0.340\times (RaNu)^{0.36\pm0.01}$ $(Pr = 3.57)$. The scaling exponents of these power laws fits are the same within experimental uncertainty but the magnitudes decrease with $Pr$.

\subsection{Local temperature fluctuations\label{SubSec_Temp_fluc}}

\begin{figure}
\centering
\includegraphics[width=\textwidth]{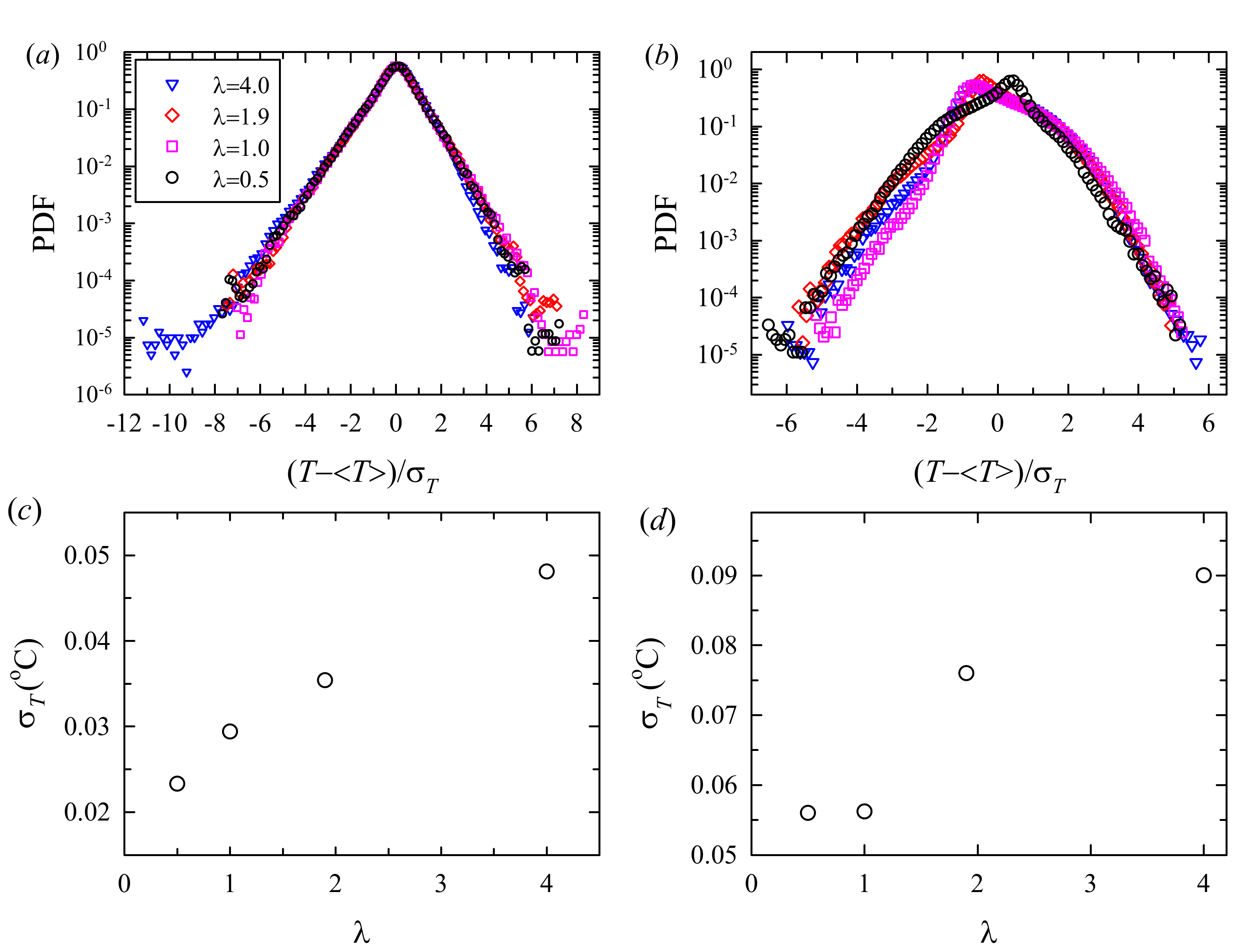}
\caption{\label{fig_PDF_localTemp} (Colour online) Probability density functions (PDFs) of the standardised temperature measured ($a$) at the cell centre and ($b$) near sidewall for different values of $\lambda$ ($Ra=2.1\times 10^{10}$ and $Pr=23.34$). The root-mean-square temperatures as a function of $\lambda$ at the cell centre and sidewall are plotted in ($c$) and ($d$), respectively.}
\end{figure}

Since Regime II is a transitional state, we focus on temperature fluctuations measured in Regime III only. To simplify the analysis, we don't take into account the effects of $Pr$ and limit the analysis to measurements done in FC770 only ($Pr$=23.34). The results may shed light upon the mechanism leading to the change of the heat transport scaling in rough cells with different roughness geometries in Regime III. 

The probability density function (PDF) of the standardised temperature $(T-\langle T \rangle)/\sigma_T$ measured at $Ra=2.1\times10^{11}$ at four different values of $\lambda$ at the cell centre and sidewall are displayed in figure \ref{fig_PDF_localTemp} ($a$) and ($b$) respectively, where $\sigma_T$ is the rms temperature that is shown as a function of $\lambda$ in figure \ref{fig_PDF_localTemp} ($c$) and ($d$) at the cell centre and sidewall. It is seen that the PDFs at different $\lambda$ overlap onto each other very nicely at the cell centre, suggesting that the bulk turbulence shares the similar dynamics at different values of $\lambda$.  A notable feature seen in  figure \ref{fig_PDF_localTemp} ($a$) is that the PDFs are significantly skewed toward the negative side, i.e. more cold plumes are detected in the cell centre than hot ones. This feature hasn't been observed either by us or reported by others in smooth cells with the same boundary conditions, i.e. constant temperature at the top boundary and constant heat flux at the bottom boundary. Thus, the reason for this phenomenon remains unknown to us at present. At sidewall, the shapes of the PDFs remain approximately invariant with $\lambda$ except the case with $\lambda=0.5$. For measurements with $\lambda=0.5$ the thermistor was so-positioned such that the upwelling hot plumes of the LSC were detected, while for the other cases the downwelling cold plumes of the LSC were detected. The positive tails of the PDFs overlap well onto each other while the negative tails, especially those larger than $3\sigma_T$, change with $\lambda$. The rms temperature at the cell centre increases considerably with $\lambda$ as can be seen from figure \ref{fig_PDF_localTemp}($c$), and that at the sidewall first remains almost the same for $\lambda$=0.5 and 1.0 and then increases with $\lambda$ as well. As the large temperature fluctuations are caused by thermal plumes, this observation suggests a change of the plume dynamics with increasing $\lambda$.

\begin{figure}
\centering
\includegraphics[width=\textwidth]{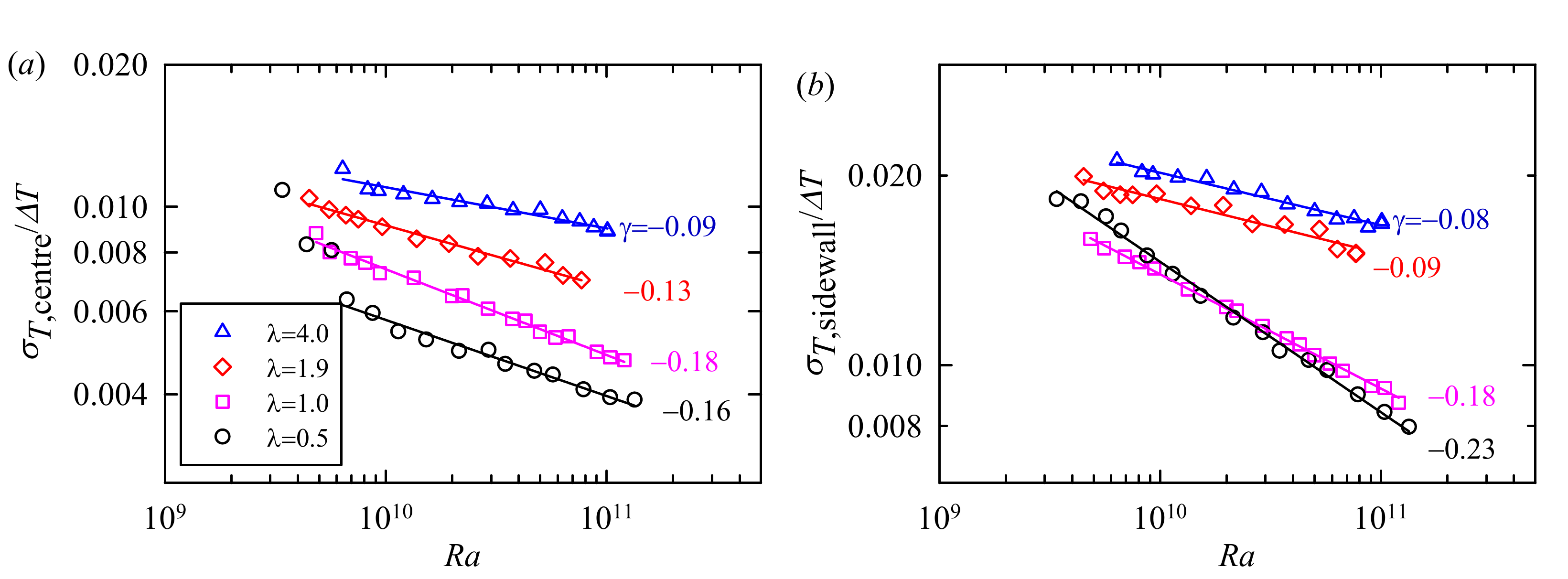}
\caption{\label{fig_rms}(Colour online) $\lambda$-dependence of the normalised root-mean-square temperature as a function of $Ra$ at ($a$) the cell centre and ($b$) the sidewall. The solid lines are power law fits to the individual data set with the scaling exponents displayed nearby each data set. The data are measured at $Pr=23.34$.}
\end{figure}

The Rayleigh number dependence of normalised rms temperatures $\sigma_{T}/\Delta T$ for different values of $\lambda$ measured at the cell centre and near the sidewall are shown in figure \ref{fig_rms}($a$) and ($b$), respectively. With increasing $\lambda$, $\sigma_T/\Delta T$ enhances considerably. The changes lie not only in the magnitudes but also the scaling exponents with $Ra$. The scaling exponents of the power law fits, e.g. $\sigma_T/\Delta T=C\times Ra^{\gamma}$, are displayed nearby each data set in figure \ref{fig_rms}($a$) and ($b$) and the scaling laws of each data set are summarised in table \ref{tab:table1}. The scaling exponent increases from -0.16 to -0.09 with $\lambda$ increased from 0.5 to 4.0 in the cell centre.  For comparison, \citet{Wei2016JFM} obtained an exponent of -0.17 in the centre of a smooth cell for a comparable $Pr$ ($Pr$ = 12.3). With a possible $Pr$-dependence of the temperature fluctuation unknown at the present, the experimental results suggest that the $Ra$-dependence of the temperature fluctuation at the cell centre becomes stronger with increasing $\lambda$. Near the sidewall, the scaling exponent increases from -0.23 to -0.08 in the $\lambda$ range covered in the experiments, suggesting a weakened dependence of temperature fluctuation on $\lambda$ at the sidewall. To the best of our knowledge,  there has been no report on the smooth-cell value for this exponent in the similar $Pr$ range near sidewall.

\begin{figure}
\centering
\includegraphics[width=\textwidth]{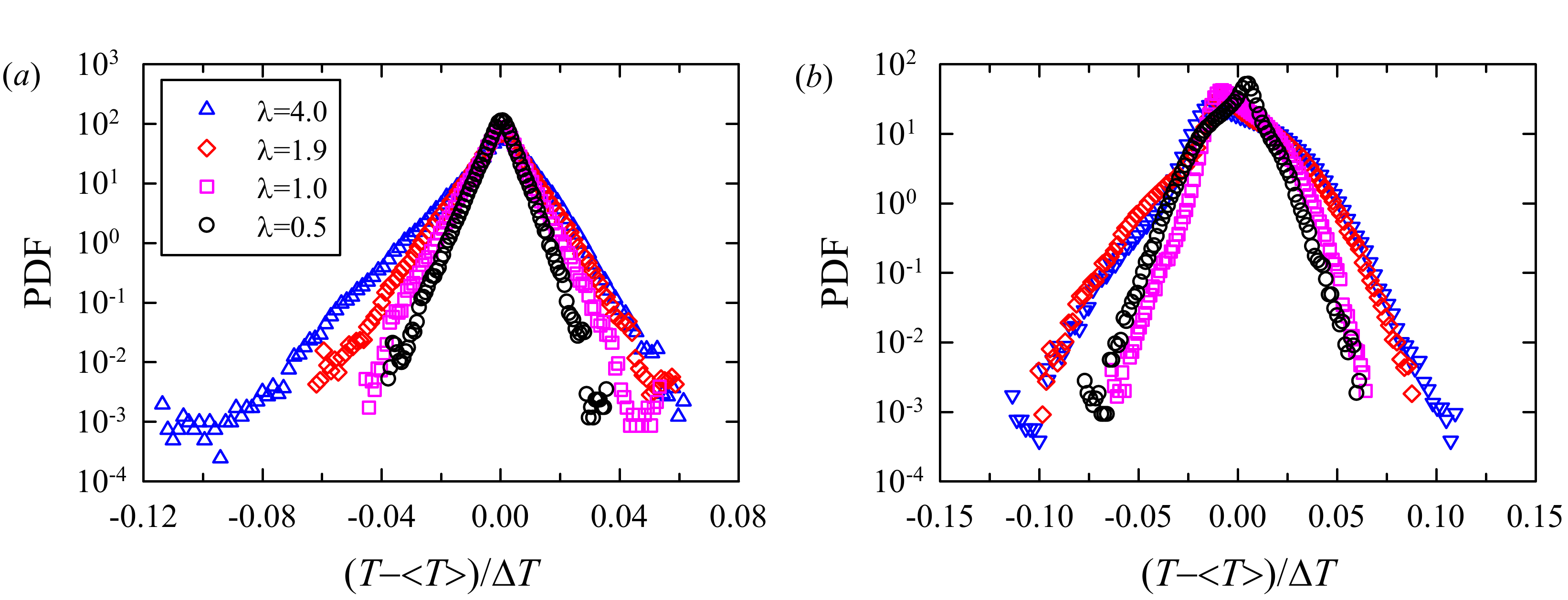}
\caption{\label{fig_sigma_T} (Colour online) Probability density function of the  temperature deviation from the mean $T-\langle T\rangle$ normalised by the applied temperature between the top and bottom plate $\Delta T$ at ($a$) the cell centre and ($b$) the sidewall ($Ra=2.1\times 10^{10}$, $Pr=23.34$). }
\end{figure}

\begin{figure}
\centering
\includegraphics[width=\textwidth]{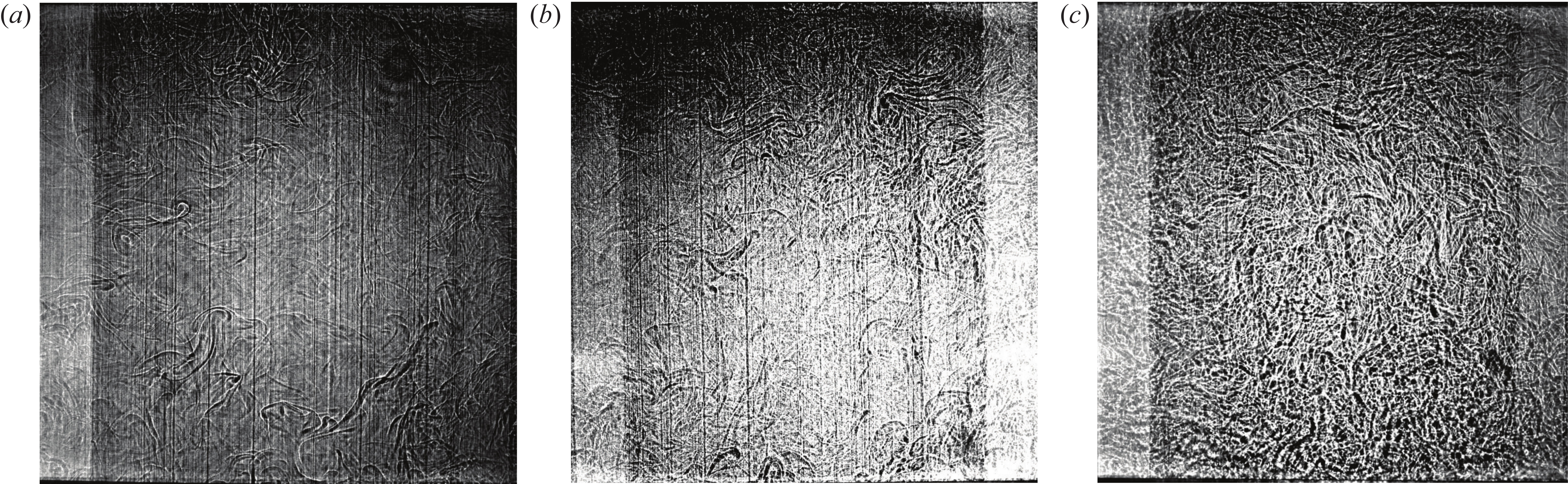}
\caption{\label{fig_shadow} Shadowgraph images of the instantaneous flow fields in a rough cell with ($a$) $\lambda$=0.0 ($b$) $\lambda=1.0$ and ($c$) $\lambda=4.0$. The images were taken at $Ra=1.22\times 10^{11}$ and $Pr=23.34$.}
\end{figure}

To gain further insight into the dynamics of temperature fluctuations with increasing $\lambda$, we examine the PDFs of the normalised temperature deviation from the mean $(T-\langle T \rangle)/ \Delta T$ for different values of $\lambda$ at $Ra=2.1\times 10^{10}$ at the cell centre and the sidewall respectively, as displayed in figure \ref{fig_sigma_T}($a$) and ($b$). The figure reveals that large events are becoming more probable with increasing $\lambda$ at the same $Ra$, i.e. hotter and colder plumes, suggesting that the effects of changing roughness geometry are to modify the dynamics of thermal plumes and make them more coherent and energetic (therefore decay slower). This is also supported by the shadowgraph visualisation of the flow field, such as those presented in figure \ref{fig_shadow}. These images are taken at $Ra=1.22\times 10^{11}$ and $Pr=$23.34. It is seen that, with increasing $\lambda$, both the density and the morphology of the thermal plumes change drastically. In the smooth cell there is a well-defined LSC with few thermal plumes passing through the cell centre (figure \ref{fig_shadow}($a$)). In the rough cell with $\lambda=1.0$, more thermal plumes are passing through the cell centre (figure \ref{fig_shadow}($b$)). In the rough cell with $\lambda=4.0$, thermal plumes occupy almost all the space in the convection cell (figure \ref{fig_shadow}($c$)). Furthermore, the plumes are now packed (or clustered) much more densely. This will result in their slower decay through thermal diffusion, which explains why it is now much more probable to find large temperature excursions in the cell centre (figure \ref{fig_sigma_T}($a$)).

\begin{figure}
\centering
\includegraphics[width=\textwidth]{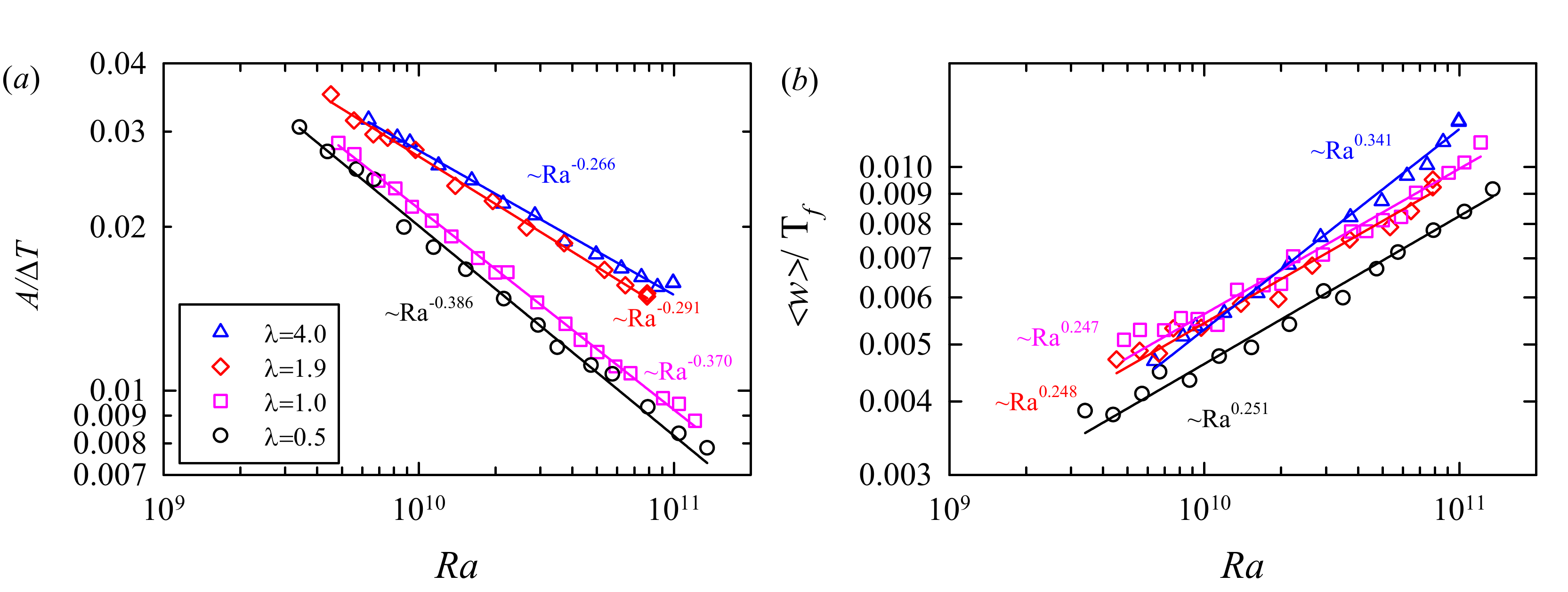}
\caption{\label{fig_plume} (Colour online) 
($a$) The time averaged plume amplitude $\langle A \rangle$ normalised by $\Delta T$ and ($b$) the time averaged plume width $\langle w\rangle$ normalised by the large-scale circulation turnover time $T_f$ as a function of $Ra$ extracted from the temperature time series measured at the sidewall at $Pr=23.34$. The scaling of the power law fits are displayed nearby each data set.}
\end{figure}

To quantitatively understand the change of the dynamics of thermal plume, we extract time series of thermal plume amplitude $A$ and width $w$ using the time-resolved temperature data measured near sidewall. The detailed plume extraction method can be found in \citet{Zhou2016PRF}. The normalised time-averaged plume amplitude $\langle A\rangle/\Delta T$ and normalised time-averaged plume width $\langle w\rangle /\tau_0$ are shown in figure \ref{fig_plume}($a$) and ($b$) respectively. Figure \ref{fig_plume}($a$) reveals that with the increase of $\lambda$, the plume amplitude $A/\Delta T$ increases. The solid lines are power law fits to the data at different values of $\lambda$ with the scaling exponents listed nearby each data set. The scaling exponent of plume amplitude increases from -0.386 to -0.266 with $\lambda$ increasing from 0.5 to 4.0. In contrast, the $Ra-$dependence of the plume width remains almost the same with a scaling exponent around 0.25 for $\lambda$ smaller than 4.0. It should be noted that the scaling exponent of plume width for $\lambda=4.0$ increases considerably (0.341) when compared with other values of $\lambda$, which may be correlated with the drastically changed plume dynamics and morphology in rough cell with $\lambda=4.0$. As the thermal plumes are main heat carriers in turbulent RBC, the experiments suggest that the change of plume dynamics may lead to the change of the scaling law of heat transport for different values of $\lambda$. 
 
\subsection{A proper parameter characterising the system: $\lambda$ vs number density of roughness elements \label{SubSec_con_para}}

When varying $\lambda$ the number density $n$ of roughness elements also varies, i.e. $n\simeq\pi/4\Gamma^2\lambda^2(H/h)^2$. In fact, the two parameters have one-to-one correspondence provided that the ratio of the roughness element height $h$ to the cell height $H$ and $\Gamma$ are fixed, which is the case for the present study. To figure out whether $\lambda$ is the most relevant parameter that characterises the system, we carried out another experiment, in which $\lambda$ was kept constant at 4.0 and the height of the roughness elements was halved. Thus the number density of the roughness elements is increased by a factor of 4. 

\begin{figure}
\begin{center}
\includegraphics[width=\textwidth]{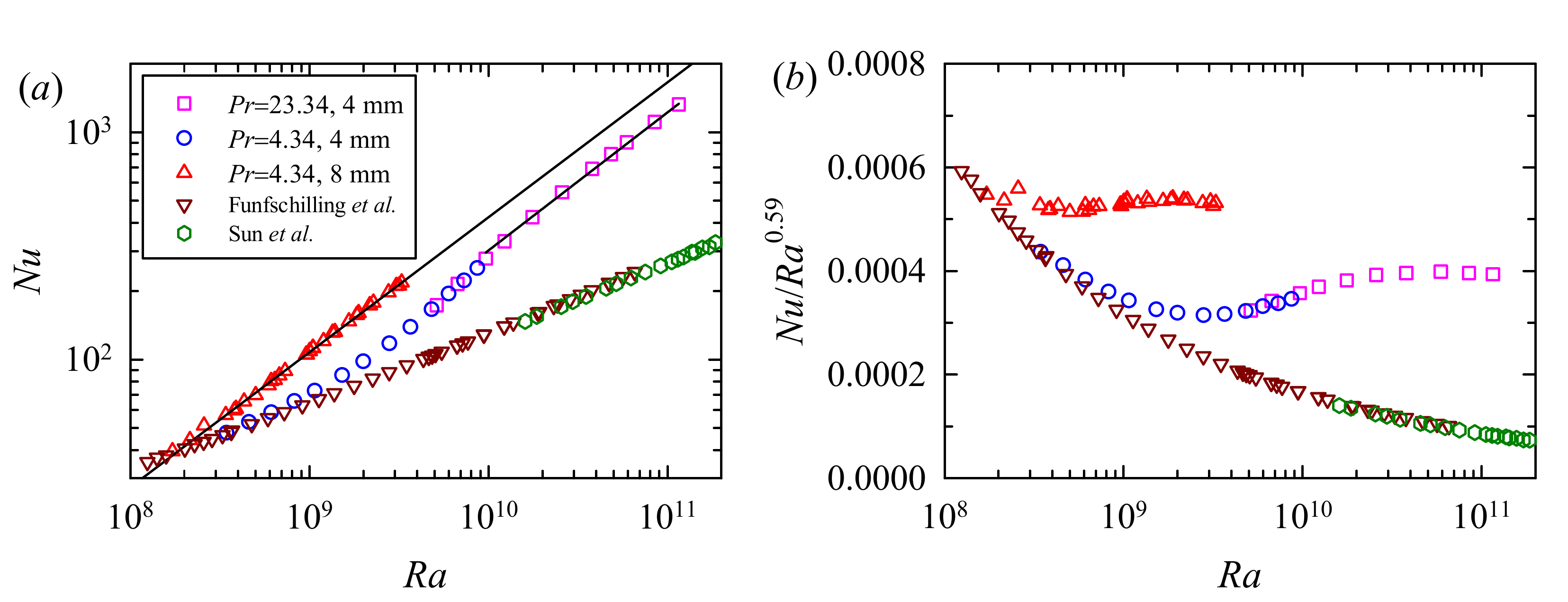}
\caption{\label{fig_Nu4} (Colour online) ($a$) $Nu$ as a function of $Ra$ and $Pr$ at fixed $\lambda=4.0$ measured in rough cells with roughness height $h$ of 4 mm and 8 mm. The down-pointing triangles are from \citet{Denis2005JFM} and the hexagons are from \citet{Sun2005JFM} in smooth cells. The solid lines are power law fits to the respective data set. The power law fitting in cell with $h=$ 8 mm is extrapolated to $Ra=2\times 10^{11}$. ($b$) Compensated plot of the same data in ($a$).}
\end{center}
\end{figure} 

\begin{figure}
\centering
\includegraphics[width=0.65\textwidth]{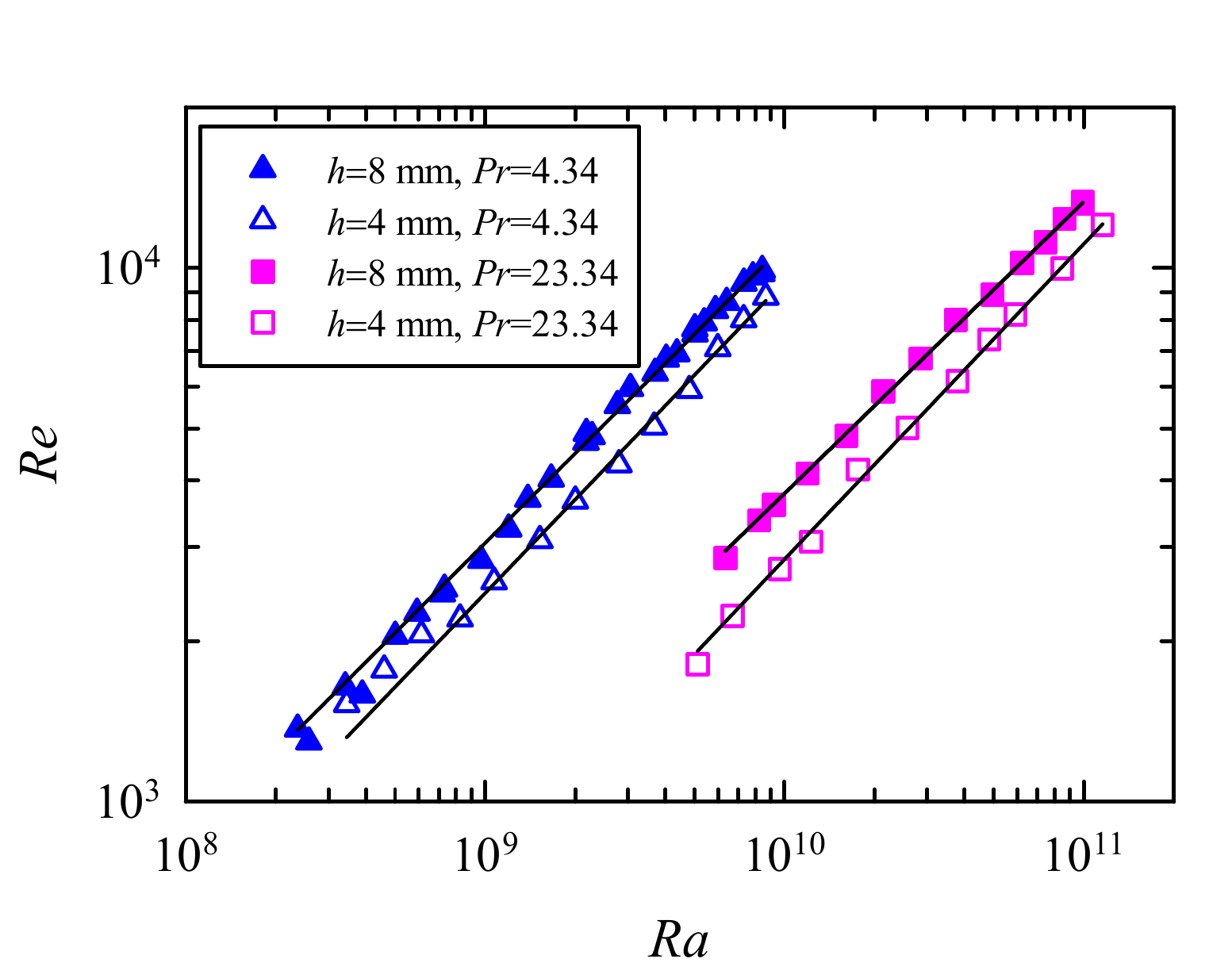}
\caption{\label{fig_Re4}(Colour online) $Re$ as a function of $Ra$ in rough cells with $h=$ 8 mm (solid symbols)  and 4 mm (open symbols) at $Pr=$ 4.34 (triangles) and 23.34 (squares), respectively. In both cells $\lambda=4.0$. The solid lines are power fits to each data set. For $Pr=4.34$, $Re=0.028R^{0.56}$ ($h=8$ mm) $Re=0.014R^{0.58}$ ($h=$4 mm); and for $Pr=23.34$ $Re=0.013R^{0.55}$ ($h=8$ mm) $Re=0.0035R^{0.59}$ ($h=$4 mm). }
\end{figure} 

We show in figure \ref{fig_Nu4}($a$) $Nu$ as a function of $Ra$ for measurements in $\lambda$ = 4.0 rough cells with $h=8$ mm and $h=4$ mm together with the same smooth cell data which have be shown in figure \ref{fig_Regime_I_II} and figure \ref{fig1}. The estimated viscous boundary layer thickness at the highest $Ra$ reached in the experiment is 4.8 mm. Thus, the measurements at $Pr=23.34$ and $h=4$ mm are in Regime II. A scaling exponent $\alpha$ = 0.60 is observed in rough cell with $h=4$ mm in Regime II, which is consistent with 0.59 observed in Regime II of rough cell with $h =$ 8 mm, which is indicated by the two parallel lines in figure \ref{fig_Nu4}. The similar $Ra$-dependence of $Nu$ in Regime II in rough cells with $h$ = 4 mm and 8 mm can be seen more clearly from the compensated plot in figure \ref{fig_Nu4}($b$). These observations suggest that $\lambda$ is not only a suitable parameter describing the geometry of the pyramid-shaped roughness elements, but also a suitable parameter to characterise the response of turbulent RBC over rough plates with different roughness geometries.

A noticeable feature from figure \ref{fig_Nu4}($a$) is that the measured $Nu$ in the rough cell with $h =$ 8 mm is larger than that in the rough cell with $h =$ 4 mm in Regime II despite the fact that they have similar $Ra$-dependence. This may be explained by the measured Reynolds number based on the circulation time of the LSC as plotted in figure \ref{fig_Re4}. It is seen that $Re$ in the rough cell with $h= 8$ mm is larger than that in the rough cell with $h=$4 mm, which is true for both $Pr =$ 4.34 and $Pr=$ 23.34. The result suggests that a faster LSC can transport heat more efficiently cross the convection cell. 

\section{Conclusion and outlook\label{Sec_con}}

We studied systematically in this paper the effects of roughness geometry on the global heat transport, the flow dynamics and  the local temperature fluctuations in turbulent Rayleigh-B\'enard convection over rough plates. The experiments were carried out in rough cells with pyramid-shaped roughness elements directly machined on the top and bottom plates. We defined a parameter $\lambda$ (the height of the roughness element over its base width) to characterise the geometry of the roughness element and to analyse the observed scaling behaviour of $Nu(Ra)$ and $Re(Ra, Pr)$, where $Ra$ and $Pr$ are respectively the Rayleigh number and Prandtl number, $Nu$ and $Re$ are respectively the Nusselt number and Reynolds number. We varied $\lambda$ from 0.5 to 4.0 by keeping the roughness height at a constant of 8 mm and changing the base width from 16 mm to 2 mm. In addition, a rough cell with $h =$4 mm was used. The experiments covered Rayleigh number $7.5\times 10^{5}\leq Ra \leq 1.31\times 10^{11}$ and $Pr$ from 3.57 to 23.34.

It is found that the heat transport is enhanced significantly by roughness elements, and the enhancement depends strongly on the roughness geometry. It is also found that the heat transport scaling, i.e. $Nu\sim Ra^{\alpha}$, may be classified in to three regimes in turbulent RBC over rough plates. In Regime I, the system is in a dynamically smooth state. The heat transport scaling in a rough cell is the same as that in a smooth cell in this regime. In Regimes II and III heat transport enhances. The scaling exponents $\alpha$ of $Nu$ vs. $Ra$ increase from 0.36 to 0.59 with $\lambda$ increased from 0.5 to 4.0 in regime II. It increases from 0.30 to 0.50 in Regime III with $\lambda$ increased from 0.5 to 4.0. The experiment thus clearly demonstrated that the heat transport scaling in turbulent RBC could be manipulated using $\lambda$ in Regimes II and III. The transition from Regime I to Regime II may be understood in terms of the thermal boundary layer thickness becoming smaller than the roughness height with increasing $Ra$ as suggested by a number of previous studies \citep{Shen1997PRL, Du1998PRL,Ciliberto1999PRL,Roche2001PRE,Tisserand2011PoF}. The transition from Regime II to Regime III may be understood in terms of viscous boundary layer crossing, i.e. the viscous BL thickness becomes thinner than the height of the roughness elements, as shown by direct measurement of the viscous boundary layer profiles in the present study. When this happens, the thermal plumes emitted from the tip of the roughness can be directly ejected into the bulk flow.  An unexpected $Pr$ effect in turbulent RBC over rough plates was also observed: Larger heat transport enhancement was observed for larger values of $\lambda$ and larger $Pr$. This may be correlated with the strong clustering of thermal plumes as clearly seen from the shadowgraph visualization of the flow field (figure \ref{fig_shadow}).

The Reynolds number $Re$ is measured based on the turnover time of the large-scale circulation evaluated near the sidewall of the cells. Both the $Ra$ and $Pr$ dependence of $Re$ were studied for different roughness geometries, i.e. $Re=B \times Ra^{\beta} Pr ^{\epsilon}$. The scaling exponent $\beta$ increased from 0.472 to 0.551 with $\lambda$ increasing from 0.5 to 4.0. and $\epsilon$ remained at the same value around $-0.7$ except for the case of $\lambda=4.0$ ($\epsilon=-0.66$). With a limited range of $Pr$, a truly increased $\epsilon$ for $\lambda=4.0$ was not for assertion. Consistent with the observation in \citet{Wei2014JFM}, it is found that $Re$ is insensitive to the plate morphology for a fixed flux Rayleigh number, i.e $RaNu$, but now with a wider range of $Pr$ and $\lambda$ as compared with the experiment of \citet{Wei2014JFM}.

Time-resolved local temperatures at the cell centre and sidewall were measured using two small thermistors. The probability density functions (PDFs) of the standardised temperature ($T-\langle T\rangle)/\sigma_T$ in the cell centre remain almost invariant with respect to $\lambda$. The measured local temperature fluctuations at both the cell centre and sidewall increase considerably with $\lambda$. The scaling exponent of the normalised rms temperature ($\sigma_T$/$\Delta T$) with $Ra$ increases from $-0.16$ to $-0.09$ at the cell centre and from $-0.23$ to $-0.08$ at the sidewall with $\lambda$ increased from 0.5 to 4.0. By extracting thermal plume amplitude and width using a method described in \citet{Zhou2016PRF}, it is found that both the plume amplitude and width increase with $\lambda$. The scaling exponent of plume amplitude increases from -0.386 to -0.266 with $\lambda$ increasing from 0.5 to 4.0. The scaling exponent for the plume width remains at the same value around 0.25 except for the case of $\lambda=4.0$ which is -0.34. As thermal plumes are the main heat carriers in turbulent RBC, the experiments suggested that that the change of thermal plume dynamics might be a possible reason leading to the change of the scaling laws between $Nu$ and $Ra$ in Regime III.

The experiments demonstrated that the heat transport behaviour in turbulent thermal convection can be manipulated using a roughness parameter $\lambda$ that characterises the geometry of the roughness elements. It further showed that the flow dynamics and local temperature fluctuations are also altered  by roughness geometry. There are several open questions: What is the effects of $\lambda$ on the dynamics of the large-scale circulation, e.g. twisting/sloshing oscillation, flow cessation/reversal? How will the vertical mean and variance temperature profiles change with $\lambda$? 

Finally, we remark that the 0.5 scaling exponent observed for $\lambda$ = 4.0 in Regime III reminds one of the Kraichnan ultimate regime of turbulent convection \citep{Kraichnan1962PoF}. Whether Regime III for $\lambda=$ 4.0 really corresponds to the Kraichnan regime needs further investigations. The place to look would be the boundary layer, i.e. a truly turbulent boundary layer should be the ultimate test for the Kraichnan regime. A very recent DNS study suggests roughness may be used as a route to trigger the transition to the ultimate state of turbulent convection \citep{Wettlaufer2017PRL}. If this is indeed the case, then the dynamics of turbulent convection in the Kraichnan regime may be studied in a Rayleigh number range that is accessible using classical fluids like water in a table top experiment.

\section*{Acknowledgments}
We thank Yu-Hao He for help with the experiment. This work is supported by the Hong Kong Research Grants Council under grant Nos. CUHK1430115 and CUHK404513.

\bibliographystyle{jfm}

\end{document}